\documentclass[aps,twocolumn,showpacs,amsmath,amssymb,pra,superscriptaddress,floatfix,longbibliography]{revtex4-1}

\usepackage{braket}
\usepackage{amsmath}
\usepackage{amssymb}
\usepackage{mathtools}
\usepackage{graphicx}
\usepackage{dcolumn}
\usepackage{bm}
\usepackage{multirow}
\usepackage{leftidx}
\usepackage{color}
\usepackage{float}
\usepackage{amsfonts}
\usepackage{eufrak}
\usepackage[german,english]{babel}
\usepackage{soul}

\newcommand \be{\begin{equation}}
\newcommand \ee{\end{equation}}
\newcommand \bea{\begin{eqnarray}}
\newcommand \eea{\end{eqnarray}}
\newcommand \bse{\begin{subequations}}
\newcommand \ese{\end{subequations}}

\newcommand \bfu{{\bf u}}

\begin{document}
\title{
Photon recoil and laser focusing limits to Rydberg gate fidelity}
\author{F. Robicheaux}
\email{robichf@purdue.edu}
\affiliation{Department of Physics and Astronomy, Purdue University, West Lafayette,
Indiana 47907}
\affiliation{Purdue Quantum Center, Purdue University, West Lafayette,
Indiana 47907}
\affiliation{Department of Physics, University of Wisconsin-Madison,
Madison, Wisconsin 53706}
\author{T. M. Graham}
\affiliation{Department of Physics, University of Wisconsin-Madison,
Madison, Wisconsin 53706}
\author{M. Saffman}
\affiliation{Department of Physics, University of Wisconsin-Madison,
Madison, Wisconsin 53706}
\affiliation{
ColdQuanta, Inc.,  111 N. Fairchild St.,  Madison, Wisconsin, 53703
}

\date{\today}

\begin{abstract}
Limits to Rydberg gate fidelity  that arise from the entanglement of internal states of neutral
atoms with the motional degrees of freedom due to the momentum kick from
photon absorption and re-emission is quantified. This occurs when the atom
is in a superposition of internal
states but only one of these states is manipulated by visible or UV photons.
The Schr\"odinger equation
that describes this situation is presented and  two cases are explored.
In the first case, the entanglement arises because the spatial wave
function shifts due to the separation in time between excitation and
stimulated emission.
For neutral atoms in a harmonic trap, the decoherence can be expressed
within a sudden approximation when the duration of the
laser pulses are shorter than the harmonic oscillator period. In this limit, the
decoherence is given by simple analytic formulas that account for
the momentum of the photon,
the temperature of the atoms, the harmonic oscillator frequency,
and atomic mass.
In the second case, there is a reduction in gate fidelity  because the photons causing
absorption and stimulated emission are in focused beam modes. This leads to a dependence of the optically induced changes in the internal states on the center of mass atomic position. 
In the limit where the time between pulses
is short, the decoherence can be expressed as a simple analytic formula
involving the laser waist, temperature of the atoms, the trap frequency
and the atomic mass. These limits on gate fidelity are studied for the standard $\pi-2\pi-\pi$ Rydberg gate and a new protocol based on a single adiabatic  pulse with Gaussian envelope. 
\end{abstract}

\maketitle


\section{Introduction}

Neutral atom qubits with Rydberg state mediated interactions have emerged as a promising platform for scalable quantum computation and simulation\cite{Saffman2016}. Trap arrays suitable for individual control of 100 and more atomic qubits have been prepared\cite{Barredo2016,Endres2016,Kumar2018,deMello2019,Brown2019,HKim2019} and high fidelity one-qubit\cite{Xia2015,YWang2016,CSheng2018} and 
two-qubit\cite{Graham2019,Levine2019,Madjarov2020} gates 
have been demonstrated. Many protocols have been proposed and analyzed for implementing two-qubit Rydberg gates\cite{Jaksch2000,XZhang2012,Xia2013,Muller2014,Rao2014,Keating2015,Theis2016b,Petrosyan2017,Han2016,Su2016,XFShi2017,Levine2019,Mitra2020,Saffman2020}. Although detailed analyses accounting for the multilevel atomic structure and finite Rydberg lifetime have identified protocols with the potential of reaching  ${\mathcal F}>0.9999$ \cite{Theis2016b,Petrosyan2016}, the studies to date, with the notable exception of \cite{Keating2015} which included a detailed examination of motional errors, have either ignored the limits set by photon recoil or treated it only approximately\cite{Jaksch2000,Saffman2005a,Wilk2010}. 

Momentum kicks due to the absorption and emission of photons during a Rydberg pulse, as well as Rydberg-Rydberg interactions,  lead to undesired entanglement between qubits encoded in hyperfine states and the atomic center of mass motion. After tracing out the motional state, the remaining entanglement in the qubit basis is reduced, which sets a limit on the gate fidelity\cite{Nielsen00}. The degree of infidelity  depends on several parameters including the magnitude of the photon momentum, the temporal extent of the Rydberg pulse sequence, the initial motional state of the atoms, the characteristic vibrational frequencies of the trap holding the atom, and whether or not the trap is turned off or left on during the Rydberg gate. 

In this paper we present a rigorous analysis of the fidelity limits set by these  effects for two representative $C_Z$ gate protocols: the standard $\pi - 2\pi - \pi$ pulse sequence\cite{Jaksch2000} and a  new,  simple implementation of a $C_Z$ gate that uses only a single Gaussian shaped adiabatic pulse applied simultaneously to both atoms. In both cases the gate infidelity scales with the change in atomic position $x$ during the gate, relative to the size of the initial center of mass wavefunction $\delta x$. For an atom that is prepared in the motional ground state of the trap, $\delta x\sim 1/\nu^{1/2}$ with $\nu$ the trap frequency so the infidelity grows proportional to $\nu$. The scaling of the infidelity follows from the observation that the change in atomic position during the Rydberg gate is fractionally more significant for a well localized spatial wavefunciton, than for a less confined wavefunction.  This counterintuitive result shows that, in contrast to many atomic implementations of quantum protocols, it is not always advantageous to work deep in the Lamb-Dicke limit of tight confinement. 

While the analysis that follows is primarily concerned with the infidelity of Rydberg gates we note that single qubit gates between internal states  are susceptible to errors analogous to those analyzed here. A prime example of this is given by qubits encoded in ground and metastable, electronically excited states in alkaline earth atoms\cite{Norcia2019} or trapped ions\cite{Haffner2008}.

 The rest of the paper is structured as follows. In Sec. \ref{sec.decoherence} we describe the unwanted entanglement that arises between internal and external degrees of freedom due to photon kicks or Rydberg forces. This is followed in Sec. \ref{sec.RydbergGates} with a description of the two gate protocols to be analyzed in detail and a brief recap of the definition of Bell fidelity which we will use to characterize the gate performance. Section \ref{sec.analytical}
 presents analytical approximations for the infidelity using  a Schr\"odinger equation treatment. Section \ref{SecRes} presents numerical infidelity results for realistic experimental parameters. The numerical results are based on a full density matrix treatment and are compared with the analytical approximations. The main results are then summarized in a concluding Sec. \ref{sec.conclusions}. Additional analysis including different cases of keeping the trapping potential on or off during the gate is provided in appendices.

\section{Qubit decoherence from motional state entanglement}
\label{sec.decoherence}

The quantum state of a trapped atom is described by its motional degrees of freedom as well as quantum numbers characterizing the nuclear and electronic states. 
Qubits are typically encoded in the electronic degrees of freedom. The most commonly used approach is encoding in hyperfine states of the ground electronic configuration\cite{Saffman2010}, although also metastable electronically excited states  may be used. The total state, including all degrees of freedom, can be written as $\ket{\Psi}=\ket{\psi}_{\rm ext}\otimes\ket{\psi}_{\rm int}$ where $\ket{\psi}_{\rm int}$ is the internal state of the atom which is used to encode a qubit and $\ket{\psi}_{\rm ext}$ is the motional state. 

Momentum transfer due to photon recoil from absorption and emission induced by the laser pulses that implement gate operations, as well as forces between Rydberg excited atoms, may  change the  motional state. When changes in the motion are correlated  with the internal state the internal and external  degrees of freedom can become entangled which, in the context of Rydberg gates,  leads to decoherence of the qubit state\cite{Saffman2010,Roghani2011}. It is perhaps worth mentioning that in other settings, notably trapped ion quantum computing, controlled entanglement between internal and external degrees of freedom is in fact crucial for gate operation\cite{Cirac1995}. 

To see this explicitly consider an initial product  of the center of mass motional state   and the qubit
\begin{equation}
|\Psi_{\rm i}\rangle =  \ket{\psi_{\rm ext}}\otimes(c_0|0\rangle +c_1|1\rangle )
\label{eq.intext}
\end{equation}
where $|c_0|^2 +|c_1|^2=1$.
After a gate operation this changes to 
\begin{equation}
|\Psi_{\rm f}\rangle = c_0  \ket{\psi_{\rm ext,0}}\otimes\ket{0} +c_1 \ket{\psi_{\rm ext,1}} \otimes\ket{1}.
\label{eq.intext2}\end{equation}
The reduced density operator for the qubit after the gate is 
$$
\rho=\begin{pmatrix}
|c_0|^2 & c_0c_1^* \chi^*\\
c_0^*c_1\chi& |c_1|^2
\end{pmatrix}
$$
where
$\chi=\bra{\psi_{\rm ext,0}}\psi_{\rm ext,1}\rangle$. Changes in the motional state that are correlated with the qubit state reduce the magnitude of the coherence  $|c_0c_1^* \chi|$. When  $|\chi|=1$ but there is a phase shift it is possible to apply a correcting rotation on the qubit. When $|\chi|<1$ the error generally  cannot be repaired. In the following sections we will explicitly calculate $\chi$ for several possible gate protocols and establish realistic experimental limits on coherence and gate fidelity.

\subsection{Momentum kick for 1 atom}

This section discusses how momentum kicks to atoms
during gate pulses affect the fidelity of the gate.
The most basic example of this situation
is a 3-state atom in a one dimensional
harmonic trap with the absorbed and emitted photons parallel to the allowed
motion. In a typical implementation 
 a laser may cause a transition from the
qubit state $|1\rangle$ but leave state $|0\rangle$ untouched.
We will limit the treatment to the case where the atom is
excited from state $|1\rangle$ to Rydberg
state $|R\rangle$ and then de-excited
as part of the gate. For simplicity, the discussion below is
in the wave function picture but all of the results in Sec.~\ref{SecRes} 
are obtained using density matrices.

A reduction in gate fidelity occurs because the laser pulses create
entanglement between the internal states of the atom and its center of
mass degrees of freedom, which has been pointed out in several 
previous works\cite{Jaksch2000,Saffman2005a,WLi2013b,Keating2015}.
Consider the case where
the atom starts in a separable wave function of the form of
Eq.~(\ref{eq.intext}).
A short laser pulse of duration $\delta t$ excites state
$|1\rangle$ to state $|R\rangle$
and in the process
gives the atom a momentum kick $\hbar K$. After a delay 
$\tau$ much shorter than the vibrational period of the atom 
trap, a second laser pulse in the same direction de-excites state
$|R\rangle$ to state $|1\rangle $ giving the atom a momentum kick $-\hbar K$
from the stimulated emission. Although the second momentum kick undoes
the change in momentum from the first, the spatial wave function for
state $|1\rangle $ will {\it not}
be the same as that for the state $|0\rangle $ because there will be a
change in position, $\delta x=\hbar K\tau /M$,
and an extra phase accumulation due to the change in
kinetic energy during $\tau$. The final wave function can be written as
\begin{equation}
|\Psi_{\rm f}\rangle =\psi_{\rm f,0}(x)c_0|0\rangle +c_1[\psi_{\rm f,1}(x)|1\rangle +\psi_{\rm f,R}(x)|R\rangle ]
\end{equation}
where the $c_j$ are as  before, the $\psi_{\rm f,0}$ is normalized to one,
and the integral of the sum $|\psi_{\rm f,1}|^2+|\psi_{\rm f,R}|^2$ is normalized to one.
Typically, the norm in state $|1\rangle $ is  much larger than that in
state $|R\rangle $.

The entanglement between the spatial and internal degrees of freedom results in
decoherence in the internal states. A measure of the decoherence is
\begin{equation}\label{EqEps}
\varepsilon=1-|\chi |
\end{equation}
where
\begin{equation}\label{EqChi}
\chi = \int_{-\infty}^\infty \psi_{\rm f,0}^*(x)\psi_{\rm f,1}(x)dx
\end{equation}
which has a magnitude
$0\leq |\chi |\leq 1$.
When $|\chi |=1$, no entanglement occurred between the spatial and
internal states and there is only an overall phase difference between
$\psi_{\rm f,0}(x)$ and $\psi_{\rm f,1}(x)$. This is the desired outcome of the
gate pulses.
There is decoherence for the case $|\chi |<1$ which
can occur even when the norm of $\psi_{\rm f,1}$ is unity
because the laser kicks can cause this part of the wave function
to evolve into a different region of Hilbert space.
From the process described in the previous paragraph, there will
be both a magnitude less than one and a complex phase for $\chi$.
See the appendices,
Sec.~\ref{SecApp} for a derivation
of $\chi$ for four different excitation styles.
We give both  the magnitude and phase for $\chi$, but the projection having
a norm less than one is more problematic because the change in
phase can be compensated for.

\subsection{Momentum transfer for 2 atoms}\label{SecKickrr}

Although not due to photon kicks from absorption or emission, there is a
source of decoherence due to transient population with double Rydberg
character\cite{Keating2015}. 
This entanglement between the internal states of the atom and the
atom motion arises because the Rydberg-Rydberg interaction depends on
the separation of the atoms leading to an impulse along the line
connecting the atoms, but only when the atoms start in the $|11\rangle$
state.
We will only consider the case where the duration of the gate pulses
is much smaller than the trap period. In this limit, the Rydberg-Rydberg
interaction leads to an impulse due to the force between the atoms
when they are both excited.
This term only enters the $\psi_{\rm f,11}$ part of the wave function
and
consists of a phase accumulation $\phi = -V_{RR}\tau_{RR} /\hbar$
where
\begin{equation}
\tau_{RR}=\int_{-\infty}^\infty P_{RR}(t)dt
\end{equation}
is effectively the time spent in the double Rydberg
state and $P_{RR}$ is the probability both
atoms are in the Rydberg state when starting in the $|11\rangle$ state.

Defining the $y$-direction as pointing from atom 1 to 2, the 
spatial dependence of the extra phase
accumulation can be approximated as $\phi =6{\sf B}(y_2-y_1)\tau_{RR}/r_{12}$
where the $y_j$ are displacements from the respective trap centers,
with $r_{12}$ the distance between the center of the
traps, and $\hbar {\sf B}=
V_{RR}(r_{12})$; this assumes a $1/r_{12}^6$
dependence on the Rydberg-Rydberg interaction and the displacements are
small compared to the trap separation. Outside of an overall,
irrelevant phase factor, the projection of the $|00\rangle$,
$|01\rangle$, or $|10\rangle$ components on that of $|11\rangle$ are
the same and is
\begin{eqnarray}\label{EqRydKick}
\langle\psi_{\rm f,00}|\psi_{\rm f,11}\rangle
&=&\langle\psi_{\rm f,00}|e^{i\phi}|\psi_{\rm f,00}\rangle
\simeq \langle 1 +i\phi -\frac{1}{2}\phi^2\rangle\nonumber\\
&=&1-\frac{36{\sf B}^2k_BT_{\rm
eff}\tau_{RR}^2}{M\omega_\perp^2r_{12}^2}
\end{eqnarray}
where $\psi_{\rm f,ii'}$ are the spatial components of the wave function
Eq.~(\ref{EqPsif2}),
$\omega_\perp$
is the trap frequency in the $y$-direction, and $T_{\rm eff}$
is defined in Eq.~(\ref{EqTeff}). We used the identity
$\langle M\omega_\perp^2 y^2\rangle = k_BT_{\rm eff}$ to generalize
Eq.~(\ref{EqRydKick}) to a thermal distribution.

\section{Fidelity analysis of Rydberg gates}
\label{sec.RydbergGates}

Our main objective is to understand the effect of photon momentum kicks
on the fidelity of Rydberg gate operations. The magnitude of the effect depends on the
gate protocol used as well as  atomic and laser parameters.  By elucidating
the role of photon momentum kicks for prototypical  examples the trends for other
cases may be apparent.

A diagonal phase  gate has the general
form, apart from an irrelevant global phase,  of
\begin{equation}
    C_\phi={\rm diag}(1,e^{\imath\phi_{01}},e^{\imath\phi_{10}},e^{\imath\phi_{11}})
\end{equation}
The requirement which $C_\phi$ must satisfy for preparation of fully 
entangled states is 
\begin{equation}
    \phi_{01}+\phi_{10}+\phi_{11}=n \pi
        \label{eq.Cphi}
\end{equation} with $n$ an odd integer. The choice $\phi_{01}=\phi_{10}=\phi_{11}=\pi$ gives a $C_Z$ gate in the standard form of 
\footnote{The textbook $C_Z$ gate is usually written as $ C_Z={\rm diag}(1,1,1,-1)$. This can be implemented by coupling state $\ket{0}$ to the Rydberg state. We will use Eq. (\ref{eq.standardCZ}) as the definition of $C_Z$ in this paper. }
\begin{equation}
     C_Z={\rm diag}(1,-1,-1,-1).
\label{eq.standardCZ}
\end{equation}
Gate protocols that satisfy (\ref{eq.Cphi}) can be converted into standard  form by applying single qubit rotations. We will assume that these can be done perfectly so that any protocol which satisifies  (\ref{eq.Cphi}) will be considered  a perfect gate implementation.

\subsection{Rydberg gate protocols}

\begin{figure}[!t]
\includegraphics[width=7.cm]{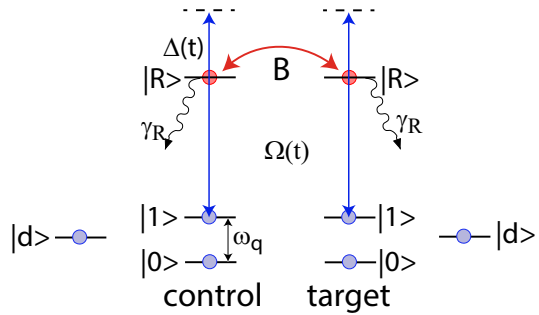}
\caption{(color online) Energy level structure of neutral atom qubits with ground states $\ket{0},\ket{1}$. Rydberg states $\ket{R}$ interact with strength $\sf B$. Rydberg states are excited with Rabi frequency $\Omega(t)$ at detuning $\Delta(t)$. Level $\ket{d}$ is an uncoupled state that accumulates spontaneous emission from $\ket{R}$ which has lifetime $\tau_R=1/\gamma_R$.
  }
\label{fig.states1}
\end{figure}

The atomic level structure is shown in Fig. \ref{fig.states1}. Atoms, each with stable ground states $\ket{0}, \ket{1}$,  are individually trapped in  harmonic potentials. State $\ket{1}$ is optically coupled to Rydberg state $\ket{R}$ while off-resonant excitation of $\ket{0}$ is assumed negligible. The small contribution to gate infidelity from this off-resonant coupling has been considered in \cite{XZhang2012,Theis2016b}. Rydberg excitation is typically implemented as a one- or two-photon process. In the latter case the photons can be sufficiently detuned from an intermediate level that the additional photon scattering is negligible. The two-photon excitation has the advantage that using a counterpropagating geometry the momentum transfer is substantially smaller than for a one-photon implementation.

\begin{figure}[!t]
\includegraphics[width=8.7cm]{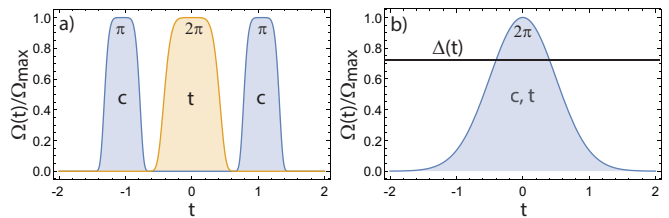}
\caption{(color online) Rydberg gate protocols. a) $\pi - 2\pi - \pi$ gate\cite{Jaksch2000} with the pulses marked c(t) applied to the control (target) qubits. b) Adiabatic gate with the same pulse applied to both qubits simultaneously with constant detuning. 
   }
\label{fig.protocols}
\end{figure}

The two protocols we analyze in detail are shown in Fig. \ref{fig.protocols}. 
The first is the  $\pi-2\pi-\pi$ protocol, the fidelity of which using constant amplitude pulses   has been analyzed in detail in \cite{XZhang2012}. Neglecting the motional effects considered here, the gate fidelity is limited by atomic structure parameters to less than ${\mathcal F}\sim 0.998$. Higher fidelity reaching ${\mathcal F} > 0.9999$ can be achieved using shaped pulses\cite{Theis2016b}. The gap between the control atom $\pi$ pulses plays a dominant role in the sensitivity of this protocol to motional decoherence\cite{Wilk2010}.

The second protocol we consider consists of a single adiabatic pulse applied simultaneously to both atoms with a  Gaussian amplitude profile and constant detuning as shown in Fig. \ref{fig.protocols}b.
This gate  is similar in concept to the Rydberg dressing gate\cite{Keating2015}, and other adiabatic gate protocols\cite{Mitra2020,Saffman2020}, but has a simplified implementation requiring only a single pulse with a constant laser detuning.

The entanglement mechanism of a single adiabatic pulse can be understood from analysis of the coherent part of the atomic dynamics depicted in Fig. \ref{fig.states1}. Consider first the simplified one-atom problem described by
$$
\hat{H}_1 = \hbar\begin{pmatrix}
0&\Omega^*/2\\
\Omega/2&-\Delta
\end{pmatrix}
$$
in the basis $(\ket{1},\ket{R})$. Here we have ignored the far detuned coupling  of $\ket{0}$ to the Rydberg state.
$\hat{H}_1$ has eigenvalues 
$\hbar\lambda_{1,2}=\hbar (-\Delta\pm\sqrt{\Delta^2+|\Omega|^2})/2$ and eigenvectors $\bfu_{1,2}$.
At the beginning of the Gaussian pulse  $\Omega(0)=0$ and the state is $\ket{\psi}=\ket{1}=\begin{pmatrix}1\\0\end{pmatrix}=\bfu_1.$ As long as the evolution is adiabatic $\ket{\psi(t)}=\bfu_1$ at all times so the atom returns to the ground state independent of the length of the pulse. This imparts robustness with respect to the amplitude of $\Omega(t)$. 

The adiabatic condition is $|\frac{d\theta}{dt}|\ll|\lambda_2-\lambda_1|=\sqrt{\Delta^2+|\Omega|^2}$ with $\theta$ the mixing angle.
The dynamical phase acquired by the state during an adiabatic 
pulse of duration $T$ is $\phi_{01}=\phi_{10}=\int_0^T dt\, \lambda_1(t)
=\int_0^T dt\, (\Delta(t)+\sqrt{\Delta^2(t)+|\Omega(t)|^2})/2$. Setting $\Delta$ constant and $\Omega(t)=\Omega_{\rm max}\left(e^{-(t-T/2)^2/\sigma^2}-e^{-T^2/4\sigma^2}\right)/\left(1-e^{-T^2/4\sigma^2}\right)$ the phase can be found numerically as a function of $\Delta, \Omega_{\rm max}$, and $\sigma.$

When the qubits are symmetrically excited  the Hamiltonian in the two-atom symmetric basis $(\ket{11},\frac{\ket{1R}+\ket{R1}}{\sqrt2},\ket{RR})$ is
$$
\hat{H}_2 = \hbar\begin{pmatrix}
0&\Omega^*/\sqrt2&0\\
 \Omega/\sqrt2&-\Delta&\Omega^*/\sqrt2\\
0&\Omega/\sqrt2&-2\Delta + {\sf B}
\end{pmatrix}.
$$
Here $\Omega$ is the one atom Rabi frequency coupling states $\ket{1}$ and $\ket{R}$. 
Diagonalizing $\hat{H}_2$ we obtain eigenvalues $\hbar\lambda_{1,2,3}$ and 
eigenvectors $\bfu_{1,2,3}.$
Explicit expressions for the eigenvalues are given in \cite{Saffman2016}.
As for the case of a single Rydberg coupled atom the dynamical phase $\phi_{11}$ is found from integrating the relevant eigenvalue over the duration of the pulse. The result depends on $\Delta, \Omega_{\rm max}, \sigma$, and $\sf B$. For any value of $\sf B$, which is fixed by the choice of Rydberg states and the interatomic spacing ${\bf r}_{12}$, we have two free parameters $\Delta/\Omega_{\rm max}$ and $\sigma$ which can be chosen to satisfy the condition on the phases for an entangling $C_Z$ gate.

Examples of valid gate parameters for different interaction strengths are given in Sec. \ref{sec.adiabatic}. As we 
will show the adiabatic gate can achieve high fidelity for a wide range of interaction strengths including the limit of strong blockade when $|{\sf B}|\gg \Omega_{\rm max}$ as well as the opposite limit of $|{\sf B}|\ll \Omega_{\rm max}$. In the latter case the doubly excited state $\ket{RR}$ is populated which leads to additional motional errors from Rydberg - Rydberg forces.

\subsection{Bell state fidelity}

We will quantify the fidelity of the $C_Z$ gate by calculating the fidelity of the 
Bell state  $|B\rangle = (|00\rangle + |11\rangle )/\sqrt{2}$ which can be 
prepared by a perfect gate operation.  
The sequence starts with
the separable state $\ket{11}$. Applying the Hadamard gate to both
qubits gives
\begin{equation}
|\rm in\rangle \equiv
{\sf H}_1{\sf H}_2|11\rangle =\frac{1}{2}(|00\rangle - |01\rangle
-|10\rangle + |11\rangle ).
\end{equation}
Applying the Rydberg $C_Z$ gate (defined to
be ${\rm diag}(C_Z) = [1,-1,-1,-1]$) followed by a Hadamard gate on qubit 2
gives the Bell state, $|B\rangle$ 
under perfect operation. In what follows, we will
evaluate the implementation
of the $C_Z$ gate by starting in a separable density matrix
\begin{equation}\label{EqBeta2}
\hat{\rho}_i =
|{\rm in}\rangle\langle{\rm in}| \rho_s ({\bf r}_1,{\bf r}_2;{\bf r}_1',
{\bf r}_2')
\end{equation}
where the $\rho_s$ contains the spatial information about the two atom system
and could be an eigenstate or a thermal state
of the two atoms. We will then solve the density matrix equation for the
$C_Z$ gate
\begin{equation}\label{EqDenMat}
\frac{d\hat{\rho}(t)}{dt} =\frac{1}{i\hbar}[\hat{H}(t),\hat{\rho}(t)]
+\hat{\cal L}(\hat{\rho})
\end{equation}
where $\hat{H}(t)$ is the time-dependent Hamiltonian which will include
the kick from photon absorption or emission. In the calculations, we
will simulate the case where the qubit $|1\rangle$ can be excited to a
Rydberg state $|R\rangle$. 
For the two-atom Hamiltonian we use 
$\hat H={\hat H}_{1}\otimes {\hat I} + {\hat I}\otimes {\hat H}_{1} +
{\hat H}_2$ where ${\hat H}_1$ is a one-atom operator and ${\hat H}_2$ describes the Rydberg interaction.

As a basic example, plane wave light propagating
in the $x$-direction will give a one atom Hamiltonian
\begin{eqnarray}\label{EqHam1}
\hat{H}_1&=&  [H_{C}
-\hbar\Delta |R\rangle\langle R|\nonumber\\
&\null & +\frac{\hbar\Omega}{2}\left( e^{iKx}
|R\rangle\langle 1| +e^{-iKx}
|1\rangle\langle R|\right) ]
\end{eqnarray}
where  $H_{C}$
represents the kinetic energy and confining potential for
each atom, $\Delta$ is the detuning of the laser connecting states $|1\rangle$
and $|R\rangle$, $K$ is the photon wave number, and $\Omega$ is
the Rabi frequency. In addition to the one atom Hamiltonian, there is a  Rydberg-Rydberg interaction
\begin{equation}
\hat{H}_2 = \hbar {\sf B} |RR\rangle\langle RR|
\end{equation}
where $\hbar {\sf B}$ is the energy shift of the pair Rydberg state.
The Rydberg state can decay by spontaneous photon emission or by
absorption or emission of
black-body photons. If the Rydberg state decays,
the time to return to the ground state manifold
can be large compared to the gate duration
and there are many more hyperfine ground states than the qubit pair. Thus,
we will use the pessimistic approximation that all of the radiative loss goes to
states outside of the qubit pair or the Rydberg state. This population can
be assigned to the single dark state $|d\rangle$ giving
\begin{equation}
\hat{\cal L}(\hat{\rho})=\Gamma\sum_{j=1}^2\left(|d\rangle\langle R_j|
\hat{\rho}|R_j\rangle\langle d|-\frac{1}{2}|R_j\rangle\langle R_j|\hat{\rho}
-\frac{1}{2}\hat{\rho}|R_j\rangle\langle R_j|\right)
\end{equation}
where $j$ labels the two atoms.

Various gate protocols have different time dependence in the coupling $\Omega(t)$,
and the detuning $\Delta(t)$. In some cases, the confining potential 
$H_{C,j}$ can depend on time and on the internal state of the atom.

In practice, we solve the density matrix equations using a basis set of
harmonic oscillator states at angular frequency
$\omega$ for the center of mass motion; in cases where more than 1D motion
is important, we will distinguish between angular frequencies of the
center of mass motion parallel, $\omega_\parallel$, or
perpendicular, $\omega_\perp$, to the beam propagation direction. This gives
\begin{eqnarray}
\hat{\rho} &=&\sum
|i_1i_2\rangle\langle i'_1i'_2|\psi_{v_1}({\bf r}_1)\psi_{v_2}({\bf r}_2)
\psi^*_{v_1'}({\bf r}_1')\psi^*_{v_2'}({\bf r}_2')\nonumber \\
&\null &\times \rho_{i_1v_1i_2v_2i'_1v'_1i'_2v'_2}
\end{eqnarray}
where $|i\rangle$ are the internal states $|0\rangle$, $|1\rangle$, $|R\rangle$,
or $|d\rangle$ and the $v$ indicate the vibrational quantum numbers where
each $v_j$ could be 1, 2, or 3 indices depending on the spatial dimension
of the simulation. The number of vibrational states needed for convergence
depends on the temperature in the simulation (more states are needed
as $k_BT/(\hbar\omega )$ increases) and the photon momentum (more states
are needed as $K$ increases).

The reduced density matrix is defined as
\begin{equation}
\rho_{i_1i_2i'_1i'_2}=\sum_{v_1v_2}\rho_{i_1v_1i_2v_2i'_1v_1i'_2v_2}
\end{equation}
where we will not use a special symbol for the reduced density matrix
since it has 4 indices instead of the 8 for the full density matrix.
The reduced density matrix is used in the calculation of the gate fidelity.

After solving the time dependent density matrix equation, Eq.~(\ref{EqDenMat}),
for the $C_Z$ gate, we add a phase to state $|1\rangle$ and then apply
a Hadamard gate to qubit 2. The phase is chosen to maximize the Bell
state fidelity. We assume the phase on state $|1\rangle$ and the Hadamard
gate can be perfectly implemented irrespective of the vibrational state.
The Bell state fidelity is defined as \cite{Sackett2000}
\begin{equation}\label{EqFidDef}
{\cal F}=\frac{\rho_{0000}+\rho_{1111}}{2}
+|\rho_{0011}|\le 1.
\end{equation}
The $C_Z$ gate protocol is designed to make $\cal F$ as large as possible. The requirement for scalable quantum computation depends on the error correction scheme\cite{Terhal2015}, but ${\cal F}>0.99$ is generally considered a minimum fidelity below which very large numbers of qubits are required.   

To simplify many of the derivations, we will use a wave function
picture to track the effects from a 
gate although all of the results are ultimately
derived from the density matrix in Eq.~(\ref{EqFidDef}).
The initial state can be written as
\begin{equation}
|\Psi_i\rangle =|{\rm in}\rangle\psi_{i}({\bf r}_1,{\bf r}_2)
\end{equation}
where $|{\rm in}\rangle$ is defined in Eq.~(\ref{EqBeta2}).
A gate will cause a manipulation of the spatial part of the
wave function depending on the internal state.
Ignoring the part of the
wave function left in the states outside of the
qubit pair, the wave function after the
gate has the form
\begin{equation}\label{EqPsif2}
|\Psi_f\rangle =\frac{1}{2}\left(|00\rangle\psi_{f,00}+|01\rangle\psi_{f,01}
+|10\rangle\psi_{f,10}-|11\rangle\psi_{f,11}\right)
\end{equation}
where the $\psi_{f,ii'}$ have the effects of different interactions
between the atoms and have norm $\sim 1$. 

Applying the Hadamard gate to
qubit 2 gives the final state
\begin{eqnarray}\label{EqPsif2H}
|\Psi\rangle &=&\frac{1}{\sqrt{2}}( |00\rangle\frac{\psi_{f,00}+
\psi_{f,01}}{2}+|11\rangle\frac{\psi_{f,11}+
\psi_{f,10}}{2}\nonumber\\
&+&|01\rangle\frac{\psi_{f,00}-
\psi_{f,11}}{2}+|10\rangle\frac{-\psi_{f,11}+
\psi_{f,10}}{2} ).
\end{eqnarray}
From this, taking $|\Psi\rangle\langle\Psi |$ and integrating over the
spatial coordinates gives the components of the reduced density matrix
\begin{eqnarray}\label{EqRhoDef}
\rho_{0000} &=&
\frac{1}{8}\left(\langle \psi_{f,00}|+\langle\psi_{f,01}|\right)\left(|\psi_{f,00}\rangle+|\psi_{f,01}
\rangle\right) \nonumber\\
\rho_{1111} &=&
\frac{1}{8}\left( \langle \psi_{f,11}|+\langle\psi_{f,10}|\right)\left(|\psi_{f,11}\rangle+|\psi_{f,10}
\rangle \right)\nonumber\\
\rho_{0011} &=&
\frac{1}{8}\left(\langle \psi_{f,00}|+\langle\psi_{f,01}|\right)\left(|\psi_{f,11}\rangle+|\psi_{f,10}
\rangle \right)
\end{eqnarray}

\section{Analytical estimates of motional gate infidelity}
\label{sec.analytical}

This section provides an analytical treatment of two different
gate protocols to give an idea of the parameters that determine the contribution
of photon momentum to the gate infidelity, $1-{\cal F}$. This derivation
often
will be based on the Schr\"odinger equation since the other effects are
small. We will then average over possible spatial states to account for
cases where the atom is not in a motional eigenstate. These analytical results
are used for interpretation; for the example results below, we always solve
the density matrix equations, Eq.~(\ref{EqDenMat}).

\subsection{Bell fidelity for $\pi - 2\pi -\pi$ $C_Z$ gate}

To
see how  entanglement between the internal qubit state and the atomic center of mass motion affects the Bell fidelity, the 
$\pi - 2 \pi -\pi$ Rydberg blockade
gate is instructive. In this case, a first pulse
excites qubit 1 to the Rydberg state if it is in state $|1\rangle$,
a second pulse
excites qubit 2 from state $|1\rangle$ to the Rydberg state and back
to state $|1 \rangle$ if qubit 1 is {\it not} in the Rydberg state,
and a last pulse de-excites qubit 1 from the Rydberg state back to
$|1\rangle$. We will consider three contributions to infidelity:
axial momentum kicks, laser focusing, and radiative losses.

\subsubsection{Contribution from axial momentum}

To derive the effect from axial momentum, the effects on
the $\psi_{f,ii'}$ from the different kicks must be
determined. The
$\psi_{f,00}$ has no kick from momentum absorption,
the $\psi_{f,10}$ has a kick on qubit 1 and
no kick on qubit 2,
the $\psi_{f,01}$ has a kick on qubit 2 and no kick on
qubit 1, and $\psi_{f,11}$ has the same kick on qubit 1 as for
$\psi_{f,10}$. All of the far off resonant qubits get a
small adiabatic kick (for example, qubit 1 in $\psi_{f,00}$
and qubit 2 in $\psi_{f,11}$).
The effect of the kick on qubit 1 can be approximated using the
results in Sec.~\ref{SSSecChi2} while that on qubit 2 for
$\psi_{f,01}$ can be approximated
using the results in Sec.~\ref{SSSecChi1}. The adiabatic kicks
can be calculated using the results in
Sec.~\ref{SSSecChiAD}. Even when the excitation to the Rydberg state
is through a nearly resonant 2 photon transition, the population out
of the ground state is small and each pulse length is small compared
to the duration of the gate. This means the adiabatic kicks
can be ignored compared to the momentum kicks.

We will write the unkicked wave function as $\psi_0$ and the
kicked wave function as $\psi_{k,1}$ for the kick on atom
1 and $\psi_{k,2}$ for the kick on atom 2. From the previous paragraph,
the different wave functions can be written as
\begin{eqnarray}
\psi_{f,00}&=&\psi_0({\bf r}_1)\psi_0({\bf r}_2)\nonumber\\
\psi_{f,01}&=&\psi_0({\bf r}_1)\psi_{k,2}({\bf r}_2)\nonumber\\
\psi_{f,10}&=&\psi_{k,1}({\bf r}_1)\psi_0({\bf r}_2)\nonumber\\
\psi_{f,11}&=&\psi_{k,1}({\bf r}_1)\psi_0({\bf r}_2)
\end{eqnarray}
The projections needed to determine the fidelity are
\begin{equation}
1-|\langle\psi_{k,i}|\psi_0\rangle|=\varepsilon^{(i)}
\end{equation}
with  $\varepsilon^{(1)}$  from Sec.~\ref{SSSecChi2},
Eq.~(\ref{EqEps2}),
while $\varepsilon^{(2)}$ is derived in
Sec.~\ref{SSSecChi1}, Eq.~(\ref{EqEps1}).

We use these terms in the wavefunction in Eq.~(\ref{EqRhoDef})
to find
\begin{eqnarray}\label{EqRhoG}
\rho_{0000} &=&
\frac{1}{4}(2 - \varepsilon^{(2)})\nonumber\\
\rho_{1111} &=&
\frac{1}{2}\nonumber\\
\rho_{0011} &=&
\frac{1}{4}(1-\varepsilon^{(1)}) (2+\varepsilon^{(2)})
\end{eqnarray}
assuming that the phase errors from the kicks are corrected. 
Putting in the analytic forms for the $\varepsilon$
and dropping terms of order $\varepsilon^2$,
the Bell infidelity
is
\begin{equation}\label{EqFid1}
1-{\cal F}=\frac{\varepsilon^{(1)}}{2}+\frac{3\varepsilon^{(2)}}{8}= \frac{K^2k_BT_{\rm eff}}{2M}\left(\frac{\tau_1^2}{2}+
\frac{3\tau_2^2}{8}\right)
\end{equation}
where $T_{\rm eff}$ is from Eq.~(\ref{EqTeff}), $\tau_1$ is the time
between the two $\pi$ pulses, and $\tau_2=\delta t/2$ with $\delta t$
the duration of the $2\pi$ pulse.
The time between $\pi$ pulses is several times
larger than the duration of the $2\pi$ pulse which
suggests $\varepsilon^{(1)}\gg\varepsilon^{(2)}$.
When the temperature is much larger than the quantized energy spacings
of the atom trap,  $T_{\rm eff}\simeq T$ and
Eq.~(\ref{EqFid1}) is identical to the
``Doppler dephasing" term in Ref.~\cite{Graham2019} Supplementary Materials
which differs by a factor of 2 from that in Ref.~\cite{Wilk2010}.

\subsection{Contribution from laser focusing}

The finite spatial extent of an atom's position combined with
spatial variation of the laser intensity due to focusing leads
to infidelity in the gate. In this section, we will only treat
this effect. There are several trends to consider.
As the laser waist decreases, the spatial variation of the intensity
increases across the spatial extent of the atom position leading
to decreased fidelity. The fidelity decreases with increasing
temperature due to the increasing spatial extent. The fidelity
increases as the trap frequency increases due to the decreasing
spatial extent.

To obtain the infidelity due to laser focusing, the projection
derived in Sec.~\ref{SSecLasFoc} will be used. The decreased
norm in each term can be found by noting that in
$\psi_{f,00}$ neither atom is excited, $\psi_{f,10}$
atom 1 is excited, $\psi_{f,01}$ atom 2 is excited,
and $\psi_{f,11}$ has no effect since the $2\pi$ pulse changes
the sign of the $\psi_{11}$ term {\it not} excited to
$\psi_{R1}$ in the first $\pi$ pulse. Unlike
the previous section, the main effect is the change in the norm
of the state associated with the transition to the Rydberg
state and back. In this case, the
different wave functions can be written as
\begin{eqnarray}
\psi_{f,00}&=&\psi_{i,00}\nonumber\\
\psi_{f,01}&=&(1-\varepsilon^{(G)})\psi_{i,00}\nonumber\\
\psi_{f,10}&=&(1-\varepsilon^{(G)})\psi_{i,00}\nonumber\\
\psi_{f,11}&=&\psi_{i,00}
\end{eqnarray}

Substituting
the projections into Eq.~(\ref{EqRhoG}), gives
\begin{equation}
\rho_{0000}=\rho_{1111}=|\rho_{0011}|=
\frac{1}{2}-\frac{\varepsilon^{(G)}}{2}
+\frac{(\varepsilon^{(G)})^2}{8}
\end{equation}
where $\varepsilon^{(G)}$ is defined in Eq.~(\ref{EqEpsG}).
Dropping all terms involving $\varepsilon^2$
the Bell state fidelity, Eq.~(\ref{EqFidDef}), gives
\begin{equation}\label{EqFidG}
1-{\cal F}=\varepsilon^{(G)}.
\end{equation}
If the excitation is due to two photon absorption, the
$1/w_0^2$ is the sum of the squares of the inverse waists
and the $1/x_R^2$ is the sum of the squares of the inverse
Rayleigh ranges.

\subsection{Contribution from radiative losses}

The contribution to the infidelity due to radiative losses can
be computed from the form of Eq.~(\ref{EqRhoDef}) by considering
the decrease in magnitude in each part of the wave function.
We will take the most pessimistic interpretation that radiative
losses due to spontaneous emission or blackbody radiation
lead to transitions outside of the qubit states on the time
scale relevant to the gate.
Assuming the Rydberg lifetime is much shorter than the gate
duration, the change in norm leads to 
\begin{eqnarray}
\psi_{f,00}&=&\psi_{i,00}\nonumber\\
\psi_{f,01}&=&\left( 1-\frac{\Gamma\tau_2}{2} \right)\psi_{i,00}\nonumber\\
\psi_{f,10}&=&\left( 1-\frac{\Gamma\tau_1}{2} \right)\psi_{i,00}\nonumber\\
\psi_{f,11}&=&\left( 1-\frac{\Gamma\tau_1}{2} \right)\psi_{i,00}
\end{eqnarray}
Using Eqs.~(\ref{EqRhoDef}) and (\ref{EqFidDef}) and
dropping terms quadratic in $\tau$ leads to the infidelity
\begin{equation}
1-{\cal F}=\Gamma\left(\frac{\tau_1}{2}+\frac{\tau_2}{4}\right)
\end{equation}
where $\Gamma$ is the inverse of the Rydberg lifetime due
to radiative losses,
$\tau_1$ is the time
between the two $\pi$ pulses, and $\tau_2=\delta t/2$ with $\delta t$
the duration of the $2\pi$ pulse.

\subsection{Bell fidelity for adiabatic $C_Z$ gate}\label{SecAdi}

Another possible gate protocol involves exciting both qubits simultaneously with  the same laser
pulse. We will let  $\Omega (t)$ have a Gaussian envelope with fixed 
 detuning $\Delta$.
This will give a $C_Z$ if the detuning, duration, and Rydberg-Rydberg
interaction, $\hbar {\sf B}$, are chosen appropriately.
The one atom Hamiltonian is Eq.~(\ref{EqHam1}) with
\begin{equation}
\Omega (t) = \Omega_0 e^{-t^2/\delta t^2}
\end{equation}
the only time dependent part.

As in the previous section, the final wave function has the form of
Eq.~(\ref{EqPsif2}) where the $\psi_{f,ii'}$ have the effects of different
kicks. The
$\psi_{f,00}$ has no kick from the photon momentum,
the $\psi_{f,10}$ has a kick on qubit 1 and no kick on
qubit 2, the $\psi_{f,01}$ has the same kick on qubit 2 and no kick on
qubit 1, and the $\psi_{f,11}$ has a kick on both qubits. There
are small kicks from far detuned transitions which will be ignored.
The results from
Sec.~\ref{SSSecChiAD} can be used to compute the effect on $\psi_{f,10}$
and $\psi_{f,01}$. The effect on $\psi_{f,11}$ is more difficult to obtain
because it involves both qubits.

Because there are 4 electronic states involved and two different atomic
momenta, we were not able to derive an expression for the projections
involving $\psi_{f,11}$. However, we have found a formula using qualitative
arguments that gives good agreement with all of the calculations we
have performed with the full density matrix. The idea is to treat the
adiabatic pulse on the $\psi_{f,11}$ state as giving the same kick to
each atom. In analogy to Eq.~(\ref{EqTau}), we define the duration of
the kick on an individual atom as half of the Rydberg population integrated
over the pulse:
\begin{equation}\label{EqTaur}
\tau_R=\frac{1}{2}\int_{-\infty}^\infty P_{R1}(t)+P_{1R}(t)+2P_{RR}(t)dt
\end{equation}
where $P_{R1}(t)$ is the probability for atom 1 to be in the Rydberg state
and atom 2 is in state $|1\rangle$, $P_{1R}(t)$ is the reverse identification,
and $P_{RR}(t)$ is the probability that both atoms are in the Rydberg state.
For the symmetric excitation of this gate, $P_{R1}(t)=P_{1R}(t)$.
This calculation requires hardly any computer time compared to solving the full
density matrix equations including the vibrational states. This leads
to the identification
\begin{eqnarray}
\psi_{f,00}&=&\psi_0({\bf r}_1)\psi_0({\bf r}_2)\nonumber\\
\psi_{f,01}&=&\psi_0({\bf r}_1)\psi_{k,1}({\bf r}_2)\nonumber\\
\psi_{f,10}&=&\psi_{k,1}({\bf r}_1)\psi_0({\bf r}_2)\nonumber\\
\psi_{f,11}&=&\psi_{k,2}({\bf r}_1)\psi_{k,2}({\bf r}_2)
\end{eqnarray}
with the projections from Eq.~(\ref{EqEpsAd})
\begin{eqnarray}
1-|\langle\psi_0|\psi_{k,1}\rangle |&=&\varepsilon^{(ad)}(\tau_a)
\nonumber\\
1-|\langle\psi_0|\psi_{k,2}\rangle |&=&\varepsilon^{(ad)}(\tau_R)
\nonumber\\
1-|\langle\psi_{k,1}|\psi_{k,2}\rangle |&=&\varepsilon^{(ad)}(\tau_R-\tau_a)
\end{eqnarray}
where $\tau_a$ is from Eq.~(\ref{EqTau}) and $\tau_R$ is from Eq.~(\ref{EqTaur}).

Ignoring all terms quadratic in the $\varepsilon$, the projections in
Eq.~(\ref{EqRhoDef}) give:
\begin{eqnarray}
\rho_{0000}&=&\frac{1}{2} -\frac{K^2k_BT_{\rm eff}}{2M}\frac{\tau_a^2}{4}\nonumber\\
\rho_{1111}&=&\frac{1}{2} -\frac{K^2k_BT_{\rm eff}}{2M}\frac{\tau_R^2+
(\tau_R-\tau_a )^2}{4}\nonumber\\
|\rho_{0011}|&=&\frac{1}{2}-\frac{K^2k_BT_{\rm eff}}{2M}
\frac{3\tau_R^2+3\tau_a^2+(\tau_R-\tau_a )^2}{8}
\end{eqnarray}
where $\tau_a$ is from Eq.~(\ref{EqTau}) and $\tau_R$ is from Eq.~(\ref{EqTaur}).
Both of these times are computed using  only  the
internal states of the atoms.
Combining these factors, the the Bell state infidelity due to
the photon momentum kicks is
\begin{equation}\label{EqFidAd}
1-{\cal F}=\frac{K^2k_BT_{\rm eff}}{2M}\left( \frac{\tau_a^2}{2}+\frac{\tau_R^2}{2}
+\frac{(\tau_R-\tau_a )^2}{4} \right)
\end{equation}
which can be compared to the same expression for the $\pi - 2\pi -\pi$
gate of the previous section,
Eq.~(\ref{EqFid1}). Although the details are different, the gates
have the same scaling with photon kick $\hbar K$, atom mass $M$, and
effective temperature $T_{\rm eff}$. An advantage of the adiabatic gate
is that the time spent in the Rydberg state, as measured by
$\tau_a$ and $\tau_R$, can be substantially shorter than the
spacing of the $\pi$-pulses $\tau_1$ in the $\pi-2\pi-\pi$ gate for the same excitation Rabi frequency.
This gate also has the advantage that it can reach high fidelity for both large and small $\sf B$, the latter case being useful for gates acting on qubits with large spatial separations.

In addition to the axial momentum kick, the Rydberg lifetime and 
the laser focus will contribute to infidelity. It is not obvious how
to analytically account for these effects for this adiabatic gate. However,
they can be accounted for by solving the small set of density matrix equations
that do {\it not} include the effect from axial recoil.

\section{Results}\label{SecRes}

In this section, we present the results of calculations that test the
accuracy of the approximations discussed above and in the appendix
and explore the
behavior of the decoherence as a function of atomic parameters.
Parameters will be chosen to correspond to transitions in Cs.
All calculations in this section will use $M=132.91$ atomic mass units.
The excitation to the Rydberg state
occurs from counter-propagating lasers of wavelength 459 and 1038~nm
which gives an effective wavelength, $\lambda_{\rm eff}\simeq 822.9$~nm,
for the photon kick, $K = 2\pi /\lambda_{\rm eff}$. For the effects
from focusing, we will use a waist ($1/e^2$ intensity radius)  of $w_0=2$~$\mu$m for each beam.
This leads to Rayleigh ranges of 27.4 and 12.1~$\mu$m and an
effective waist $w_0=\sqrt{2}$~$\mu$m and effective Rayleigh range
of 11.1~$\mu$m in Eq.~(\ref{EqFidG}).

We will use two example Rydberg states in the calculation,
66S and 106S. The lifetime of these states,
130 and 366~$\mu$s respectively, include stimulated absorption
and emission due to blackbody radiation\cite{Beterov2009,Beterov2009b}.

\subsection{$\pi -2\pi -\pi$ $C_Z$ gate}

The first tests will involve parameters for the $\pi -2\pi -\pi$ gate
of Ref.~\cite{Graham2019}. To model the pulses, we use
\begin{eqnarray}
\Omega_1 (t) &=& \Omega_{1,\rm max}\left(e^{-(t+\tau_1/2)^6/\delta t_1^6} +
e^{-(t-\tau_1/2)^6/\delta t_1^6}\right)\label{EqOmeg1}\\
\Omega_2 (t) &=& \Omega_{2,\rm max}e^{-t^6/\delta t_2^6}
\end{eqnarray}
where the $\Omega_{j,\rm max}$ give the appropriate $\pi$ pulses for
atom 1 and $2\pi$ pulse for atom 2, $\tau_1=1.0044$~$\mu$s is
the time between pulses for atom 1, $\delta t_1=0.14$~$\mu$s is the
duration of each of the atom 1 pulses, and $\delta t_2=0.22$~$\mu$s is
the duration of the atom 2 pulse. Because of the order of magnitude
difference between $\tau_1$ and $\tau_2 =\delta t_2/2$,
the Bell state infidelity is $1-{\cal F}\simeq \varepsilon^{(1)}/2$.

\begin{figure}
\resizebox{80mm}{!}{\includegraphics{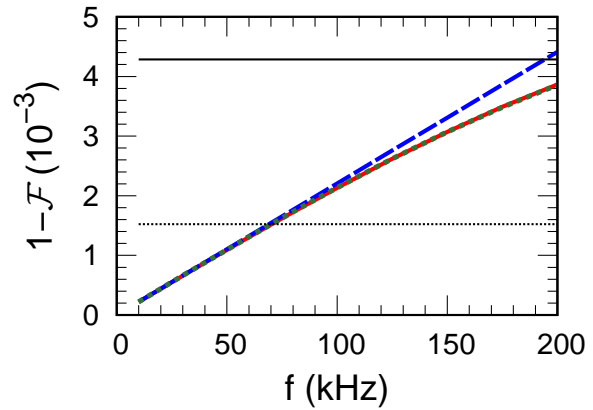}}
\caption{\label{FigMagD}(color online) 
The infidelity, $1-{\cal F}$ in Eq.~(\ref{EqFid1}), due to axial momentum
kick for
the atom in the ground state of a harmonic trap as a function
of the trap frequency $f$, using $T=0$ and
not including radiative decay of the Rydberg
state. The trap potential is on
for the full sequence, Eq.~(\ref{EqOmeg1}), for the parameters in
the text. This calculation assumes the trapping potential is magic
so that the potential is the same for all internal states.
The solid red line is the full density matrix calculation,
the long dashed blue line is from the approximation, Eq.~(\ref{EqEps1}),
and the green dotted line is the approximation that includes the
trapping potential, Eq.~(\ref{EqEpsHO}). The full density matrix
calculation and Eq.~(\ref{EqEpsHO}) are indistinguishable. The
horizontal solid black line (66S) and dotted black line (106S)
are the contribution to infidelity due to the finite lifetime
(130 and 366~$\mu$s respectively) of these Rydberg states.
}
\end{figure}

Figure~\ref{FigMagD} shows the infidelity, $1-{\cal F}$ in
Eq.~(\ref{EqFid1}), for these parameters as a function of the
trapping frequency when the atoms start in the motional ground state.
For this case, we assume a scenario where the trapping potential
is magic so that the potential is the same for all states.
The finite lifetime of the Rydberg state is {\it not included}
in this calculation, but the horizontal lines show the contribution
to the infidelity
due to the Rydberg finite lifetime for 66S (solid) and 106S (dotted) states.
The solid red line is from numerically solving the density matrix
calculation; the approximation, Eq.~(\ref{EqEpsHO}), that includes the
trapping potential is the green dotted line that overlays the exact
result. This shows that the small width of the individual $\pi$-pulses
has a negligible contribution to $\varepsilon^{(1)}$.
The blue dashed line is the exact result when the
trapping potential is turned off during the
calculation, Eq.~(\ref{EqEps1}). Because this calculation has
$\sim 1$~$\mu$s between $\pi$-pulses, it is not surprising that
Eq.~(\ref{EqEps1}) becomes noticeably different from calculations
with the trap on when $f\sim 100$~kHz. For these calculations,
convergence to better than 0.01\% was achieved using a maximum
vibrational quantum number of 10.

It is, perhaps, a surprise that there is more, not less, decoherence
as the trapping frequency increases. Intuition suggests that higher
frequencies lead to less effects from atom recoil due to the larger
energy spacing. However, because the durations are short, the size
of the energy spacings are not relevant. The main effect is from
the spatial
size of the initial state and how far the atom moves during the
sequence of laser pulses. The spatially smaller states at higher frequency
have less overlap with the unkicked states. At the higher frequencies,
the infidelity due to the axial photon kick can be larger than that from
radiative losses even when the atom is in the motional ground state.

\begin{figure}
\resizebox{80mm}{!}{\includegraphics{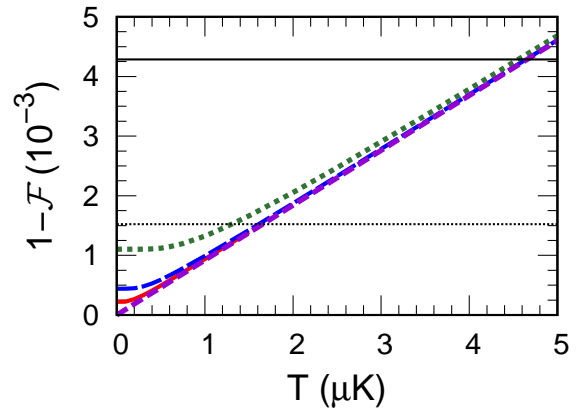}}
\caption{\label{FigMagDTh}
The infidelity, $1-{\cal F}$, due to axial momentum kicks
for the $\pi -2\pi-\pi$ gate with
the atom in a thermal distribution of a harmonic trap as a function
of the temperature, $T$. The calculation does not include
radiative decay of the Rydberg
state. The different plots are for different trap
frequencies: 10~kHz (red solid), 20~kHz (blue dashed), and 50~kHz
(green dotted). The purple
short dashed line is the high temperature approximation,
Eq.~(\ref{EqDecAsym}), with the second term in the square
brackets dropped. The approximation, Eqs.~(\ref{eqDecSh},\ref{EqEps2}), are
indistinguishable from the full calculation on this scale.
The horizontal lines are the same as Fig.~\ref{FigMagD}.
}
\end{figure}

For the laser parameters of the previous paragraph, we now compare
the approximation, Eq.~(\ref{EqEps2}), to the full calculation with the
trapping
potential for a thermal distribution. Results from calculations
with three different trap frequencies are shown in Fig.~\ref{FigMagDTh}.
These calculations include the trapping potential during the laser
manipulations.
The results from the approximation, Eq.~(\ref{EqEps2}),
are indistinguishable from the full solution on the scale shown.
As in Fig.~\ref{FigMagD}, the horizontal lines are the infidelity
due to finite Rydberg lifetime of the 66S (solid) and 106S (dotted) states.
The fidelity is 
${\cal F}>0.995$ for the temperatures plotted. The best value for
the 50~kHz trapping potential is ${\cal F}\simeq 0.999$.
The maximum number of vibrational states was used for the 20~kHz
calculation at 5~$\mu$K because the 10~kHz calculation was only
performed to 1.5~$\mu$K; the maximum number of vibrational states
for this case was $\sim 100$.

Figure~\ref{FigMagDTh} shows that the decoherence
increases linearly with the temperature once $k_BT$ is larger
than $\sim\hbar\omega_\parallel$. Also, the decoherence is nearly independent
of the frequency of the trapping potential when this condition is
satisfied because the smallest relevant length scale is the
thermal de Broglie wavelength. Since
the approximation, Eq.~(\ref{EqEps2}), is accurate, we can use it to derive
an expression for the decoherence when $\varepsilon^{(1)}$ is small:
\begin{equation}
\varepsilon^{(1)} \simeq \frac{K^2\tau_1^2k_BT_{\rm eff}}{2M}
=\frac{K^2\tau^2\hbar\omega_\parallel}{4M}{\rm coth}\left(\frac{\hbar\omega_\parallel }{2k_BT}\right)
\label{eqDecSh}
\end{equation}
where ${\rm coth}(x)=(e^x+e^{-x})/(e^x-e^{-x})$. The
limits are
\begin{eqnarray}
\varepsilon^{(1)}& \to &\frac{K^2\tau_1^2\hbar\omega_\parallel}{4M}
\left[ 1 +2e^{-\hbar\omega_\parallel /k_BT}\right]
\quad {\rm for}
\; k_BT\ll\hbar\omega_\parallel \\
&\to & \frac{K^2\tau_1^2k_BT}{2M}\left[
1+\frac{1}{12}\left(\frac{\hbar\omega_\parallel}{k_BT}
\right)^2\right]\; {\rm for}\; k_BT\gg\hbar\omega_\parallel
\label{EqDecAsym}
\end{eqnarray}
which shows the linear dependence on the temperature when
the atoms are hot compared to the quantum energies. It also shows
that the decoherence does not depend on the trap frequency at
high temperatures. Another point of comparison
is to remember that the average number of vibrational quanta is 
$\langle n\rangle =1/\{ \exp [\hbar\omega_\parallel /(k_BT)]-1\}$. When $k_BT=\hbar\omega_\parallel$,
the average vibrational quanta is $\langle n\rangle \simeq 0.58$ and the fractional
error in the high temperature form of the decoherence is 8\%. When $k_BT=2\hbar\omega_\parallel$ then $\langle n\rangle \simeq 1.54$ and the fractional error is 2\%. This illustrates
how quickly the trap frequency becomes irrelevant to the decoherence.

As a physical example, we consider the case from Ref.~\cite{Madjarov2020}
where Sr is excited from the $^3$P$_0$ metastable state to $^3$S$_1$
Rydberg states using a single 317~nm photon. The time scale for a
$2\pi$  pulse was of order 100~ns. Using their estimated
temperature of $2.5/\sqrt{10}$~$\mu$K$ = 0.8$~$\mu$K gives
$\varepsilon^{(1)}\simeq 1.5\times 10^{-4}$. However, if the gate is composed of more than
one pulse, the duration can be longer. The decoherence increases to
$\varepsilon^{(1)}\simeq 1.3\times 10^{-3}$ if  $\tau_1 =300$~ns.

\begin{figure}
\resizebox{80mm}{!}{\includegraphics{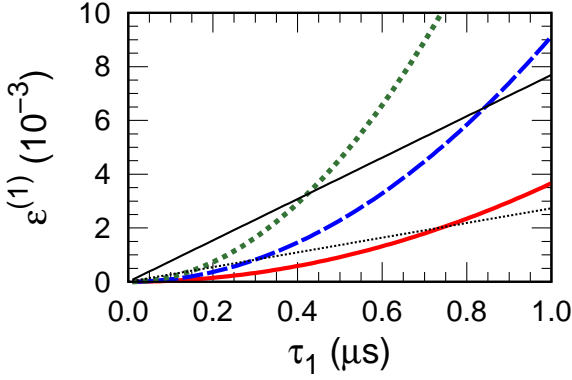}}
\caption{\label{FigMagDTau}
The projection error, $\varepsilon^{(1)}$,
due to axial momentum kick for
atoms in a thermal distribution of a harmonic trap as a function
of the time between $\pi$-pulses. All calculations are for a trap
frequency of 10~kHz using
Eq.~(\ref{eqDecSh}). The different plots are for different atom
temperatures: 2~$\mu$K (red solid), 5~$\mu$K (blue dashed), and 10~$\mu$K
(green dotted). The solid (66S) and the dotted (106S) black lines are
$\Gamma\tau_1$.
}
\end{figure}

The scale of the projection error can be compared to the decoherence
from decay of the Rydberg state. Figure~\ref{FigMagDTau} shows the
$\varepsilon^{(1)}$ for calculations in a 10~kHz trap for three different
temperatures (2, 5, and 10~$\mu$K) as a function of the time, $\tau_1$, between
the $\pi$-pulses. This emphasizes that the infidelity from
 Rydberg state  decay
scales linearly with $\tau_1$ while that from recoil is proportional to
the square of the time. For short times between pulses, the main error
will be due to the decay of the Rydberg state. For pulses with
$\tau_1\sim 1$~$\mu$s, the atoms need to be cold for the projection error
to be smaller than the decay probability.

\begin{figure}
\resizebox{80mm}{!}{\includegraphics{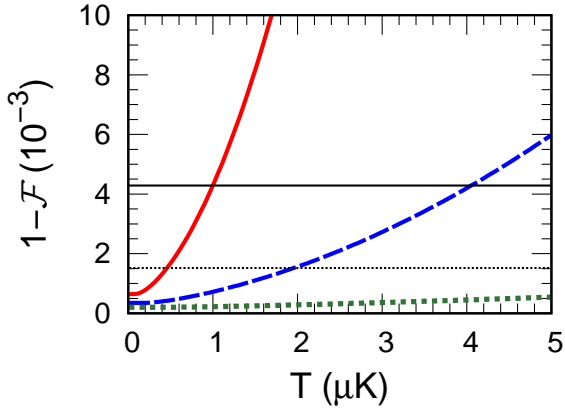}}
\caption{\label{FigMagDThFoc}
(color online) The infidelity, $1-{\cal F}$, due to laser focus, Eq.~(\ref{EqFidG}),
for the $\pi -2\pi-\pi$ gate with
the atom in a thermal distribution of a harmonic trap as a function
of the temperature, $T$. The calculation does not include
radiative decay of the Rydberg
state. The different plots are for different transverse trap
frequencies: 10~kHz (red solid), 20~kHz (blue dashed), and 50~kHz
(green dotted). This calculation does not include the axial decoherence
since it is much smaller than that for the transverse degrees of
freedom.
The horizontal lines are the same as Fig.~\ref{FigMagD}. A
transverse misalignment of $y_0=100~\rm nm$ is assumed.
}
\end{figure}

\begin{figure}
\resizebox{80mm}{!}{\includegraphics{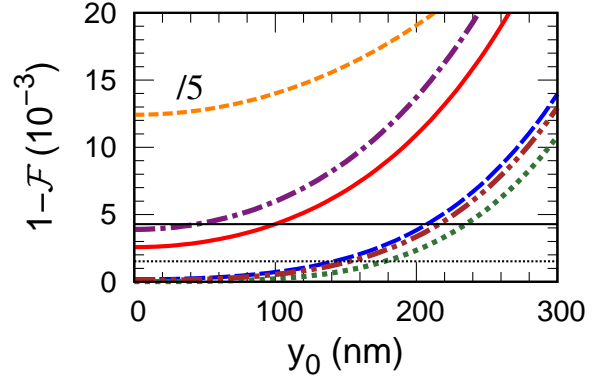}}
\caption{\label{FigMagDThFocDis}(color online) 
The infidelity, $1-{\cal F}$, due to laser focus, Eq.~(\ref{EqFidG}),
for the $\pi -2\pi-\pi$ gate with
the atom in a thermal distribution of a harmonic trap as a function
of the displacement, $Y_0$. The calculation does not include
radiative decay of the Rydberg
state. The different plots are for different transverse trap
frequencies and temperatures: 10~kHz, 1 $\mu$K (red solid), 
20~kHz, 1 $\mu$K (blue dashed), and 50~kHz, 1 $\mu$K
(green dotted), 10~kHz, 5 $\mu$K (orange short dash), 
20~kHz, 5 $\mu$K (purple dash-dot), and 50~kHz, 5 $\mu$K
(brown dash-dot-dot). The 10~kHz, 5 $\mu$K infidelity is divided by
5 to fit on the same graph.
This calculation does not include the axial decoherence
since it is much smaller than that for the transverse degrees of
freedom.
The horizontal lines are the same as Fig.~\ref{FigMagD}.
}
\end{figure}

The infidelity due to laser focusing, Eq.~(\ref{EqFidG}), does not depend
on the duration of the gate as long as the duration is much less than
the trap period. This infidelity has the strongest dependence on the
effective temperature of the mechanisms examined in this paper.
Figure~\ref{FigMagDThFoc} shows the infidelity, Eq.~(\ref{EqFidG}),
for 3 different transverse trap frequencies: 10~kHz (solid red),
20~kHz (blue dashed), and 50~kHz (green dotted). For this calculation
and that of Fig.~\ref{FigMagDThFocDis},
we assumed the terms with $x_R$ are less than 10\% of the transverse
decoherence and have
been dropped. The effective waist is $\sqrt{2}$~$\mu$m and there is
assumed misalignment of $y_0=100$~nm.
This figure illustrates how sensitive the infidelity is to
the transverse trap frequency: at high temperature, the infidelity is proportional
to $1/f_\perp^4$. This figure also shows
how high temperature exacts a high penalty because the infidelity
scales like $T^2$ in this limit.

Figure~\ref{FigMagDThFocDis} shows how the axial displacement, $y_0$, can
quickly degrade the fidelity of the $\pi - 2\pi -\pi$ gate. This graph
uses the same parameters as Fig.~\ref{FigMagDThFoc} except the temperatures
and frequencies are fixed: the curves use 10, 20, or 50~kHz and
1 or 5~$\mu$K. As with Fig.~\ref{FigMagDThFoc}, it is clear that 
larger temperatures dramatically degrade the fidelity as do
lower trap frequencies. Figure~\ref{FigMagDThFocDis} demonstrates the
importance of alignment for the gate fidelity since the infidelity,
Eq.~(\ref{EqFidG}), has a quartic dependence on $y_0$. It is interesting
that the infidelity from displacement of the focused beam is greater
than that due to finite Rydberg lifetime  for displacements larger than $\sim 200$~nm
even for the $66S$ state; this value is $\sim 1/10$ the waist of the
individual beams.

Because the infidelity is proportional to the atom temperature for
$k_BT>\hbar\omega$, the change in vibrational character due to
gate pulses should be examined. This effect can be estimated
using the heating of the atom's center of mass motion due to
the time delay between the pulses. The kick shifts the position
which leads to heating.
For reasonable gate parameters, this effect will be negligible except
for high frequency traps at the lowest temperatures.
The center of mass energy will change by the
amount
\begin{equation}\label{EqDelEkick}
\Delta E = \frac{1}{2}M\omega^2\delta x^2 = \frac{\hbar^2K^2}{2M}(\omega\tau_1 )^2
\end{equation}
which is the recoil energy times $(\omega\tau_1 )^2$. Since $\omega\tau_1$ will
be much less than one, this shows the energy will increase by a small
fraction of the recoil energy per gate. To give an idea of the size of 
the two effects, for
$\tau_1 = 1.00$~$\mu$s and $\lambda_{\rm eff}=822.9~\rm  nm$,
the change in energy from Eq.~(\ref{EqDelEkick}) is
0.42~nK per kick for a 10~kHz trap, 1.7~nK per kick
for 20~kHz, and 11~nK per kick for 50~kHz. 

As a comparison, the change
in energy that occurs when the trap is turned off for a time
$\tau \simeq \tau_1 + 4\delta t_1\simeq 1.56$~$\mu$s and
then turned back on is $\Delta E = (k_BT_{\rm eff}/2)
(\omega\tau )^2$. For this effect, the size of the
energy change increases with each time the trap is turned off and
back on. Taking $<E>=k_BT_{\rm eff}$,
The temperature after $N$ gates is $T_N=T_{\rm eff}\exp [N 
(\omega \tau_1)^2/2]$. For 100 gates, $T=T_{\rm eff}e^{0.48}=1.62
T_{\rm eff}$ for a 10~kHz trap, $e^{1.92}T_{\rm eff}=6.83T_{\rm eff}$
for a 20~kHz trap, and $e^{12}T_{\rm eff}=1.6\times 10^5T_{\rm eff}$
for a 50~kHz trap. Clearly, several approximations break down for
the 50~kHz trap before this limit is reached. Nevertheless this result indicates
that turning on and off a 50~kHz trap is not a good idea. Fortunately it is possible to design traps that present the same trapping potential for  
ground and Rydberg atoms so it is feasible to leave the trap potential on continuously\cite{SZhang2011}.  

\subsection{Adiabatic $C_Z$ gate}
\label{sec.adiabatic}

This section contains results for the adiabatic gate discussed in
Sec.~\ref{SecAdi}. Examples of parameters that lead to an effective
$C_Z$ gate are in Table~\ref{TabAdi}. The values for $\tau_1$,
Eq.~(\ref{EqTau}), result from one atom calculations and 
$\tau_R$, Eq.~(\ref{EqTaur}), result from two atom calculations
that do not include atomic recoil. The values
for $1-{\cal F}$ in this table result from calculations that do not include the
atomic recoil but do include the radiative loss from the Rydberg state.
In Table~\ref{TabAdi},
the infidelity is almost completely due to the radiative loss; the
number of photons absorbed or emitted is equal to the infidelity to the digits
given. If the radiative loss is {\it not} included, the infidelity
solely arises
from non-adiabaticity and from non-ideal choice of gate parameters
to generate a $C_Z$ gate; ignoring radiative loss, the infidelity would
be $1.8\times 10^{-5}$, $9.9\times 10^{-8}$, and $1.3\times 10^{-5}$ for the three cases shown.

\begin{table}
\caption{Parameters for the adiabatic gate. All have
$\Omega_0=2\pi\times 17$~MHz.  Parameters $\tau_1, \tau_R$ are defined in Eqs. (\ref{EqTau}), (\ref{EqTaur}). The infidelity 
$1-{\cal F}$ is that due to the pulse shapes, finite blockade, and
includes finite Rydberg lifetime, but does {\it not} include momentum kicks.
The left infidelity is for 66S and the right is for
106S. See Fig.~\ref{FigTrent} for an example
that includes momentum kicks.}
\label{TabAdi}
\centering
\begin{tabular}{c c c c c c }
\hline
Gate & $\Delta/\Omega_0$\hskip 5 pt\null & $\delta t$ 
($\mu$s)\hskip 5 pt\null & ${\sf B}$ $(10^6 s^{-1})$\hskip 5 pt\null
&$\tau_1,\tau_r$ (ns)\hskip 5 pt\null&$1-{\cal F}$ ($10^{-4}$)\\
\hline
1\hskip 5 pt\null &  -0.5000 & 0.2 &  2$\pi$ 600 &90, 63&6.3, 2.4\\
2\hskip 5 pt\null & -0.8635 & 0.2165 & 2$\pi$ 60&56, 42&3.8, 1.3\\
3\hskip 5 pt\null & -0.3000  & 0.5 & 2$\pi$ 4&357, 416&30, 11\\
\hline
\end{tabular}
\end{table}

Compared to the gate in the previous section, the resulting
duration in the Rydberg state, $\tau_{1,R}$, is over an order of magnitude
shorter for Gates 1 and 2 and a factor of $\sim 3$ smaller for Gate 3.
This should translate to a factor of $\sim 100$ (Gates 1 and 2)
or $\sim 10$ (Gate 3) improvement in fidelity due to momentum kicks
with a factor of $\sim 10$ and $\sim 3$ improvement of radiative
decay of the Rydberg state.

\begin{figure} 
\resizebox{80mm}{!}{\includegraphics{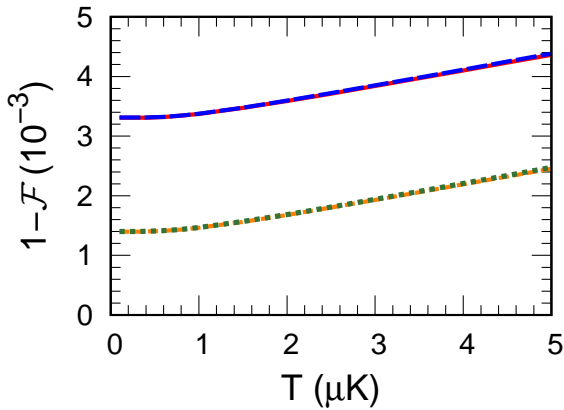}}
\caption{\label{FigTrent}
The Bell state infidelity for Gate 3, see Table~\ref{TabAdi}, as a function
of atom temperature. The trap has a frequency of 50~kHz. The solid red
line is from the full density matrix calculation and the dashed
blue line is from Eq.~(\ref{EqFidAd}) added to the intrinsic $1-{\cal F}$
from Table~\ref{TabAdi} for 66S. The orange short dash and green dotted
are the same for 106S.
}
\end{figure}

Figure~\ref{FigTrent} shows the infidelity,
$1-{\cal F}$, versus the atom temperature
for gate 3 which is the
worst gate in the table because it assumes the weakest Rydberg interaction and therefore requires a relatively  long interaction time leading to more spontaneous emission.
The calculation is performed for a trapping frequency
of 50~kHz to reduce the size of the density matrix calculation. Higher
temperature or lower frequency requires more states for convergence.
The two atom kick calculations have up to 40 vibrational states
for each atom for the 5~$\mu$K calculation
and goes to 10th order in the expansion of the exponential
$e^{\imath  K x}$.

As expected, the infidelity is a factor of $\sim 10$ smaller than
the gates of the previous section. The full results were compared to the
simple expression Eq.~(\ref{EqFidAd}), with the intrinsic gate
error of Table~\ref{TabAdi} added to it. As seen in the figure, 
there is good agreement between the full calculation and the simple analytic
results. In this temperature range, the main infidelity for this gate
is due to radiative losses in the Rydberg state. The infidelity from
the momentum kicks becomes larger than that from radiative losses
at temperatures a bit above 10~$\mu$K.

In this gate, population does transiently occupy the double Rydberg
state which means there can be additional entanglement between the
internal states and the motional character due to Rydberg - Rydberg forces as discussed in
Sec.~\ref{SecKickrr}.
Examining the projections
in Eq.~(\ref{EqRhoDef}), the definition of Bell state fidelity in
Eq.~(\ref{EqFidDef}), and the overlap factor in Eq.~(\ref{EqRydKick}),
this contributes a decrease to the fidelity of
\begin{equation}
1-{\cal F}=\frac{3}{8}[1-\langle\psi_{f,00}|\psi_{f,11}\rangle ]
= \frac{27{\sf B}^2k_BT_{\rm eff}
\tau_{RR}^2}{2M\omega_\perp^2r_{12}^2}
\end{equation}
where the parameters are defined below Eq.~(\ref{EqRydKick})
and the thermal average was taken in the expectation value of
 Eq.~(\ref{EqRydKick}). Unlike
the momentum kick from the photon, the trap frequency explicitly appears
in the infidelity. As the trap frequency increases, the infidelity
decreases as expected.

This expression was checked by solving the full density matrix equations
for Gates 2 and 3 for several temperatures
and found to be accurate at the couple percent level.
Calculations gave $\tau_{RR}$ of 31.6 ps, 8.60 ns, and 157 ns for
Gates 1, 2, and 3. For Gate 1, the separations are 2.6~$\mu$m and
5.3~$\mu$m for the 66S and 106S states while for Gate 2 the separations
are 4.2~$\mu$m and 10.5~$\mu$m and for Gate 3 are 8.0~$\mu$m and 20.0~$\mu$m.
For
$T_{\rm eff}=5$~$\mu$K and 50~kHz trap frequency, the infidelities are
(Gate 1: $9.0\times 10^{-5}, 2.2\times 10^{-5}$),
(Gate 2: $2.5\times 10^{-2}, 4.1\times 10^{-3}$), and 
(Gate 3: $1.0\times10^{-2}, 1.7\times 10^{-3}$).

\section{Conclusions}
\label{sec.conclusions}

We have derived the equations needed to analyze the decoherence
that arises from the entanglement of internal states of neutral
atoms with the motional degrees of freedom. The entanglement
occurs when the internal states of the atom are manipulated by
laser pulses. Although entangled atomic states can be prepared using collisional interactions without resorting to laser excitation, Rydberg gates have the advantage that they are fast, operate at long range, and can reach high fidelity without preparation of atoms in the motional ground state. 

Different levels of approximation were
considered above and it was shown that a sudden approximation
well describes the decoherence in this system. This approximation
gives a simple formula for the decoherence, Eq.~(\ref{EqChi1}),
in terms of the shift in position due to the photon absorption and
re-emission, $\delta x$, and an initial state length scale, $\Delta x$.
We also derived the infidelity that occurs  when the lasers that excite
the atom are focused.

The trends in the decoherence are important for future experiments,
especially in the limit of small decoherence $\varepsilon\ll 1$.
The decoherence is quadratic in the shift in position due to the
absorption and re-emission of the photons which means the
decoherence is quadratic in the separation of laser pulses, $\tau$,
and in the momentum kick from the photon, $\hbar K$. 
It is also inversely quadratic in the initial state length scale $\Delta x$;
larger $\Delta x$ leads to less decoherence and vice versa. 
For cold atoms, the initial state length scale is that for a harmonic
oscillator which is inversely proportional to the trap frequency.
This leads to the decoherence being proportional to the trap frequency
at low temperatures. At high temperatures, this length scale is
the thermal de Broglie wave length which is inversely proportional
to the temperature. This leads to the decoherence being proportional
to the temperature at high temperatures. Finally the combination that
$\delta x$ is inversely proportional to the mass and $\Delta x$ is
inversely proportional to the square root of the mass leads to the
decoherence being inversely proportional to the mass.

The direction of the trends are as might be expected with the
possible exception that the decoherence is less for small trap
frequencies (i.e. it is not necessary to get into the motional
ground state, but it is necessary to be cold). The particular power
law of the dependence might not be expected and shows where
it is important to do better. For example, if the momentum kick
is decreased by a factor of 2 by using a pair of counter-propagating
photons, the decoherence will decrease by a factor of 4.
The trends discussed above and the simple expression for
the decoherence will be a useful guide for quantum computation
applications.

For focused lasers the $\pi-2\pi-\pi$ gate infidelity increases with the deviation of the atomic position from the focal position of the  excitation beam. Inspection of Eq. (\ref{EqEpsG}) shows that there is a temperature independent error associated with misalignment of the trapping potential and the excitation laser, followed by contributions that scale as temperature  and temperature squared. The temperature dependent errors are inversely proportional to  the trap frequency to the second and fourth powers. We have not analyzed the focusing error for the adiabatic gate, but anticipate that it will be significantly smaller due to the insensitivity of the adiabatic pulse area to the laser amplitude\cite{Saffman2020}.  

The trend for the focusing error is that it is less for larger transverse trap frequencies. This is opposite to the motional infidelity which favors lower axial trap frequencies. This the case for the $\pi-2\pi-\pi$ gate and the adiabatic gate analyzed here. Similar trends are expected for other gate protocols.   The expressions derived  for the dependence of gate infidelity on trap parameters, atomic temperature, and pulse parameters will be useful for reaching high fidelity regimes needed for quantum information applications.

\begin{acknowledgments}
F.R. was supported by the National Science Foundation under Grant
No. 1804026-PHY. The work at UWM was supported by 
NSF PHY-1720220, NSF  award 2016136,
ARL-CDQI Cooperative Agreement No. W911NF-15-2-0061, 
DOE award DE-SC0019465, and DARPA Contract No. HR001120C0068. 
\end{acknowledgments}

\appendix

\section{Analytic results for 1 atom}\label{SecApp}

\subsection{Basic equations}

The derivation in this section will be based on the Schr\"odinger equation
to simplify the interpretation. All of the calculated results,
Sec.~\ref{SecRes},
were found by numerically solving the full density matrix equation,
Eq.~(\ref{EqDenMat}), except for those presented in
Figs.~\ref{FigMagDThFoc} and \ref{FigMagDThFocDis}.

To solve for the decoherence measured by $\chi$, Eq.~(\ref{EqChi}), the
full Schr\"odinger equation needs to be tracked through the
different excitation and de-excitation
steps. By solving the full Schr\"odinger equation, we automatically
include all of the effects like change in vibrational level, spatial change,
momentum change, etc. For a 3 state atom,
the three state wave function will be written
as
\begin{equation}
|\Psi (t)\rangle  =\sum\psi_{j}(x,t)|j\rangle e^{-i{\cal E}_jt/\hbar}
\end{equation}
where ${\cal E}_j$ is the energy of internal state $|j\rangle$ and
we have incorporated the $C_j$'s into the spatial function
to make the notation simpler. The spatial functions are solutions
of the equations
\begin{eqnarray}
i\hbar\frac{\partial\psi_0}{\partial t} &=& H_{C,0}\psi_0\nonumber\\
i\hbar\frac{\partial\psi_1}{\partial t} &=& H_{C,1}\psi_1+V_{1R}e^{-i\omega_{R1}t}\psi_R\nonumber\\
i\hbar\frac{\partial\psi_R}{\partial t} &=& H_{C,R}\psi_R+V_{1R}e^{i\omega_{R1}t}\psi_1
\end{eqnarray}
where the $H_{C,j}$ is the kinetic energy operator plus the trapping
potential for state $|j\rangle$,
$V_{1R}(x,t)=\hbar\Omega (t)\cos (\bar{\omega }t- K x)$
is the coupling between internal states $|1\rangle $ and $|R\rangle $ due to
the laser, $K$ is the wave number of the photon,
and $\omega_{R1}=({\cal E}_R-{\cal E}_1)/\hbar$. The
$\Omega (t)$ is the Rabi frequency for the 1-$R$ transition. The
frequency of the laser, $\bar{\omega}$, might also have time dependence
if the laser is chirped.

The rotating wave approximation should work very well for this system
since the transition frequency is large and the laser is weak. This leads
to the approximation
\begin{eqnarray}
i\hbar\frac{\partial\psi_1(x,t)}{\partial t} &=& H_1\psi_1(x,t)+
\frac{\hbar\Omega (t)}{2}e^{-i K X}\psi_R(x,t)\nonumber\\
i\hbar\frac{\partial\psi_R(x,t)}{\partial t} &=&
H_r\psi_R(x,t)+\frac{\hbar\Omega (t)}{2}e^{i K X}\psi_1(x,t)
\label{EqSchx}
\end{eqnarray}
where the $e^{\pm \imath K x}$ leads to the momentum kicks
during the absorption and stimulated emission steps.
These equations can be solved numerically using many different
techniques, leap-frog, Crank-Nicolson, etc. In cases where we numerically
solved the Schr\"odinger equation,
we used the leapfrog algorithm for the time propagation.

The situation described above has the {\it initial} $\psi_1(x,t)=C_1\psi_{\rm in}(x)$
and $\psi_R(x,t)=0$. After the pair of laser pulses, the $\psi_R(x,t)\simeq 0$.
There is some population left in state $|R\rangle $ due to Doppler shifts
of the wave function but the population is small for the cases
discussed in this manuscript.

The amplitude $\Omega (t)$ is chosen to give approximately 100\%
transition from 1-to-$R$ and from $R$-to-1. This condition can be accomplished
in a variety of ways: a single $2\pi$-pulse, two $\pi$-pulses, etc. For example
a simple form that satisfies
these requirements and models excitation to the Rydberg state
with a time delay $\tau$ before de-excitation is
\begin{equation}\label{EqOmeg}
\Omega (t)=\frac{\sqrt{\pi}}{\delta t}\left( e^{-t^2/\delta t^2}+
e^{- (t-\tau )^2/\delta t^2}\right) .
\end{equation}

\section{No trapping potential, $V_{\rm tr}=0$:}

In many experiments, the trapping potential for the atom will be dropped
during the laser excitation and de-excitation steps. In this situation,
the $H_{C,0}=H_{C,1}=H_{C,R}=P^2/(2M)$. This may seem a special case, but
in most experiments it is expected that the total duration of all
laser pulses will be much shorter than the oscillation period of the
trapping potential. We found that using the approximation $V_{\rm tr}=0$ in this case
also led to accurate results for the momentum kicks.

Equations~(\ref{EqSchx}) can be recast using the Fourier transform
\begin{equation}
\phi_j(k,t)=\frac{1}{\sqrt{2\pi}}\int_{-\infty}^\infty e^{-ikx}\psi_j(x,t)dx
\end{equation}
by multiplying from the left by $\exp (-ikx)/\sqrt{2\pi}$ and
integrating over $x$. This gives the Schr\"odinger equation for
the Fourier transforms
\begin{eqnarray}
i\hbar\frac{\partial\phi_1(k,t)}{\partial t}=E(k)\phi_1(k,t)+
\frac{\hbar\Omega (t)}{2}\phi_R(k+K,t)\nonumber\\
i\hbar\frac{\partial\phi_R(k,t)}{\partial t}=E(k)\phi_R(k,t)+
\frac{\hbar\Omega (t)}{2}\phi_1(k-K,t)
\label{eqSchk}
\end{eqnarray}
where $E(k)=\hbar^2 k^2/(2M)$ is the kinetic energy. This may appear
to be just as complicated to solve as the Schr\"odinger equation
in $x$, Eqs.~(\ref{EqSchx}), but it is actually much simpler. By writing
the second equation at $k+K$, every $k$ corresponds
to a simple 2-state system. The
$k$-space equations, Eq.~(\ref{eqSchk}), reduce to ${\cal N}_k$
pairs of equations where ${\cal N}_k$ are the number of $k$-points in
$\phi_j(k,t)$. We label the functions on a grid in $k$ using equally
spaced points $k_i = k_0 + i\cdot \delta k $ where $i =0, 1, ... {\cal N}_k-1$
as $\phi_{1,i}(t) = \phi_1(k_i,t)$ and $\phi_{R,i}(t)=\phi _R(k_i+K,t)$.
Note the $\phi_{R,i}$ are at shifted momenta compared to the $\phi_{1,i}$
This results in the set of equations
\begin{eqnarray}\label{eqSchkd}
i\hbar\frac{\partial\phi_{1,i}(t)}{\partial
t}&=&E(k_i)\phi_{1,i}(t)+\frac{\hbar\Omega (t)}{2}\phi_{R,i}(t)\nonumber\\
i\hbar\frac{\partial\phi_{R,i}(t)}{\partial
t}&=&E(k_i+K)\phi_{R,i}(t)+\frac{\hbar\Omega (t)}{2}\phi_{1,i}(t)
\end{eqnarray}
where the energy is shifted in the second equation to account for the
shift in the $\phi_{R,i}$\cite{Keating2015}.

Because these equations do not couple the $\phi$ at different $i$, a
massive simplification in the calculation of the decoherence occurs.
The final $\phi$ can be written in terms of the initial $\phi$ and
a unitary rotation that depends on, $i$:
\begin{equation}
\phi_{j,i}(t_f)=\sum_{j'} U_{jj'}(i;t_f,t_{0})\phi_{j',i}(t_{0})
\end{equation}
where $t_0$ is the initial time and $t_f$ is the final time
The simplification arises because the decoherence projection
$\chi$ only depends on the $U_{11}(i;t_f,t_0)$ projected
on the initial state. Since the overlaps do not depend on the representation
\begin{eqnarray}
\chi&=&\int_{-\infty}^\infty\phi^*_{f,0}(k)\phi_{f,1}(k)dk\nonumber\\
&\simeq &\delta k\sum_i|\phi_{\rm in,i}|^2e^{iE(k_i)(t_f-t_0)/\hbar}U_{11}(i;t_f,t_0)
\end{eqnarray}
where the initial function is $\phi_{\rm in,i}=\phi_{\rm in}(k_i)$ and we used the
fact that the $\phi_{0,i}(t)=\exp (-iE(k_i)t/\hbar)\phi_{\rm in,i}(0)$. Thus,
within the rotating wave approximation and the discretization of the
$k$-space wave functions, the decoherence overlap is
given by
\begin{equation}
\chi=\sum_i\delta k|\phi_{\rm in,i}|^2{\cal K}_{11}(i)
\end{equation}
where ${\cal K}_{11}(i)=\exp (iE(k_i)(t_f-t_0)/\hbar)U_{11}(i;t_f,t_0)$ is the
kernel for the momentum $k_i$ to start in state $|1\rangle $ at time $t_0$,
propagate forward in time to $t_f$ with the laser pulses, and then
propagate backward in time to $t_0$ with no laser pulses.

\section{Thermal distribution: $V_{\rm tr}=0$}

Using the theoretical development of the previous section, the decoherence
overlap for a thermal distribution is derived in this section. The initial system
is an incoherent sum over eigenstates, $\phi_\alpha (k)$, with the probability of each state
being $P_\alpha = \exp (-E_\alpha /[k_B T])/Z$ with
$Z = \sum_\alpha \exp (-E_\alpha /[k_B T])$. One way to treat this
is to say that the initial state is
\begin{equation}
\phi_{in,i}=\sum_\alpha\sqrt{P_\alpha}e^{i\theta_\alpha}\phi_{\alpha}(k_i)
\end{equation}
where the phases, $\theta_\alpha$ are random. This leads to the
decoherence overlap being
\begin{eqnarray}\label{EqChirho}
\chi &=&\sum_i\delta k {\cal K}_{11}(i)\sum_{\alpha\alpha'}\sqrt{P_\alpha P_{\alpha'}}\phi_{\alpha}(k_i)\phi_{\alpha'}(k_i)e^{i(\theta_\alpha-\theta_{\alpha'})}\nonumber\\
&=&\sum_i \delta k {\cal K}_{11}(i)\rho (k_i,k_i)
\end{eqnarray}
where the $k$-space density matrix is
\begin{equation}
\rho (k_i,k_i)=\sum_\alpha P_\alpha \phi_\alpha^2 (k_i)
\end{equation}
arises because all of the cross terms in $\alpha ,\alpha'$ average to 0.
Thus, for the cases where the trapping potential can be neglected
during the laser manipulations, the decoherence for a thermal distribution
is just as easy to obtain as for a particular state.

The density matrix in Eq.~(\ref{EqChirho})
depends on the temperature and the potential the atoms were trapped
in just before the laser manipulations occur. Typically, this is a
nearly harmonic potential.

The density matrix for a thermal distribution in a harmonic oscillator
can be exactly obtained for any temperature. The Wigner function
for a thermal distribution is
\begin{equation}
W(x,k) = C \exp (-[\Delta k x]^2-[\Delta x k]^2)
\end{equation}
where $\Delta k^2 = M\omega^2/(2 k_BT_{\rm eff} )$,
$\Delta x^2 = \hbar^2/(2Mk_BT_{\rm eff})$, and the $T_{\rm eff}$ is
defined in Eq.~(\ref{EqTeff}).  The  $M$ is the atomic mass, $k_B$ is Boltzmann's constant,
and $T_{\rm eff}$ is an effective temperature given below.
The density
matrix can be obtained by Fourier transform
\begin{eqnarray}
\rho \left( k+\frac{\bar{k}}{2},k-\frac{\bar{k}}{2} \right) =\int e^{-i\bar{k}x}W(x,k)dx\nonumber\\
=\frac{\Delta x}{\sqrt{\pi}}\exp\left( -[\Delta x k]^2- \frac{\bar{k}^2}{4\Delta k^2}  \right)
\end{eqnarray}
and a similar expression can be obtained for the spatial density matrix
by Fourier transforming on $k$. Although tedious, one can show that this is the exact
density matrix by using $\rho_T = C \rho_{T/2}\rho_{T/2}$.

The diagonal density matrix
is obtained by setting $\bar{k}=0$:
\begin{equation}
\rho (k,k) = \frac{\hbar}{\sqrt{2\pi Mk_BT_{\rm eff}}}\exp (-\hbar^2k^2/[2 M k_B T_{\rm eff}])
\label{eqRhok}
\end{equation}
where we have substituted for $\Delta x$. The $T_{\rm eff}$ can be
fixed by forcing the average energy from the classical form,
$\bar{E}=k_BT_{\rm eff}$, to equal the average energy from the quantum
thermal distribution,
\begin{equation}\label{EqTeff}
k_BT_{\rm eff}=\hbar\omega\left(\frac{1}{2}+\frac{1}{e^{\hbar\omega /(k_BT)}-1}   \right).
\end{equation}
which gives the limit $k_BT_{\rm eff}\to\hbar\omega /2$ at low temperatures and
$k_BT_{\rm eff}\to k_BT $ at high temperatures.

Thus, the decoherence overlap for a thermal distribution is obtained by using the
result of Eqs.~(\ref{eqRhok},\ref{EqTeff}) in Eq.~(\ref{EqChirho}).

\section{Overlaps for $V_{\rm tr}=0$}

This section derives approximate expressions for the decoherence overlap
for different excitation/de-excitation procedures. All cases assume the
center of mass coordinate is in a thermal distribution and that the
trapping potential is off during the gate. The latter condition is a
good approximation if the gate duration is short compared to the trapping
period. We note that
having the atom in the motional ground state is a thermal distribution
with $T\to 0$ which gives
$k_BT_{\rm eff}\to\hbar\omega /2$. We account for
the atom recoil during the stimulated absorption and emission process
in all of the cases.

\subsection{Short, $2\pi$ pulse, on resonance}\label{SSSecChi1}

This section derives the $\chi$ when there is a single, short pulse that
takes 100\%
of the atom population from state $|1\rangle$ to the Rydberg state and
back to state $|1\rangle$. If the atom is treated as stationary, this leads
to a $\pi$ phase shift for state $|1\rangle$.
We assume
that the duration of the laser pulse is short enough that the band width of the
laser is much larger than the Doppler width and the spacing of the
energy levels. In this limit, the shape of
the pulse is not important and we choose for it to be a flat top function:
\begin{equation}
\Omega (t) = \frac{2\pi}{\delta t}\quad 0\leq t\leq \delta t
\qquad \Omega (t) = 0\; {\rm otherwise}.
\end{equation}
For this $\Omega (t)$, Eqs.~(\ref{eqSchkd}) can be solved exactly. Defining
the parameters
\begin{eqnarray}
\delta E_i &=&E(k_i+K)-E(k_i)\nonumber\\
\bar{E}_i&=&\frac{E(k_i+K)+E(k_i)}{2}
\end{eqnarray}
the final population in the state $|1\rangle$ is $1$ minus a term of order
$(\delta E\delta t/\hbar)^4$ so it is only the phase dependence that contributes
to $\chi$. We find
\begin{equation}\label{EqK111}
{\cal K}^{(2)}_{11}(i)\simeq -e^{-i\delta E_i\tau_2 /\hbar}
\end{equation}
where we have defined $\tau_2 =\delta t/2$ is the time spent in the Rydberg state.

We can use this approximation with the thermal density matrix, Eq.~(\ref{eqRhok}),
to obtain an analytic expression for the decoherence overlap.
The energy difference is $\delta E_i=E_{\rm rec}+\hbar^2k_iK/M$
where $E_{\rm rec}=\hbar^2K^2/(2M)$ is the recoil energy of the
atom. Thus, the
overlap is reduced to the Fourier transform of
a Gaussian giving:
\begin{equation}\label{EqChi1}
\chi^{(2)}=-e^{-iE_{\rm rec}\tau_2 /\hbar}e^{-(\delta x /[2\Delta x])^2}
\end{equation}
where $\delta x=\hbar K\tau_2 /M$ is the distance an atom
shifts due to the absorption and re-emission of photons separated
in time by $\tau_2$ and
$\Delta x= \hbar/\sqrt{2k_BT_{\rm eff}M}$ is proportional to the
de Broglie wave length at the effective temperature, $T_{\rm eff}$,
in Eq.~(\ref{EqTeff}).

This form leads to a nice interpretation of the decoherence. The
phase has a $-1$ from the pulse and has an accumulation
that arises from the average change in kinetic energy of the atom
due to the photon absorption, $E_{\rm rec}$, for photons separated by $\tau_2$.
The decrease in normalization arises because the atoms shift position
due to their changed velocity over the duration $\tau_2$. At low temperatures,
the shift is compared to the spatial width of the ground state wave
packet because $k_BT_{\rm eff}\simeq \hbar\omega /2$. At higher temperatures,
the shift is compared to the coherence length of the atomic packet
which is proportional to the thermal de Broglie wave length.

If the phase is corrected, this gives a value
\begin{equation}\label{EqEps1}
\varepsilon ^{(2)} \simeq \left( \frac{\delta x}{2\Delta x}\right)^2
=\frac{K^2\tau_2^2k_BT_{\rm eff}}{2M}
\end{equation}
where the Taylor series expansion of the Gaussian in Eq.~(\ref{EqChi1})
was used because the duration is short. This result
shows that $\varepsilon$ for this case is proportional to: the square of the
duration of the pulse, the effective temperature, $T_{\rm eff}$
in Eq.~(\ref{EqTeff}), the square of the photon momentum, and the inverse of the
atom mass. This suggests which parameters can be used to suppress decoherence due
to momentum kick. For example, a one photon excitation with 319~nm photons is approximately
6.6 times worse than excitation with counter propagating 459 and 1038 nm photons.

\subsection{Two short, $\pi$ pulses, on resonance}\label{SSSecChi2}

This section derives the effect from a case like in Eq.~(\ref{EqOmeg}).
We will assume $\tau $ is much less than the period of center of mass motion.

As a first approximation, we treat the case where the $\delta t\ll \tau_1$
which allows a very simplified derivation. In this
case, we treat the excitation and de-excitation steps as
being instantaneous and can be thought of as a sudden approximation.
The final wave function is obtained from the concatenation of
three steps. The first step is the excitation from state $|1\rangle $ to
$|R\rangle $ with the momentum kick:
\begin{equation}
\phi_R(k+K) = -i\phi_{\rm in,1}(k)
\end{equation}
where the $-i$ results from the $\pi$-pulse. The second step is
the free evolution of state $|R\rangle $ for a time $\tau_1$:
\begin{equation}
\phi_R(k+K) = -i\phi_{\rm in,1}(k)e^{-iE(k+K)\tau_1/\hbar}
\end{equation}
where the phase accumulation is at the shifted momentum. The
third step is the de-excitation giving:
\begin{equation}
\phi_{f,1}(k)=(-i)^2\phi_{\rm in,1}(k)e^{-iE(k+K)\tau_1/\hbar}
\end{equation}
where the $-i$ again results from the $\pi$-pulse. This gives
the sudden approximation of the kernel:
\begin{equation}\label{EqK112}
{\cal K}_{11}^{(1)}=-e^{-i[E(k+K)-E(k)]\tau_1 /\hbar}.
\end{equation}

This is the same form as Eq.~(\ref{EqK111}) which means the $\chi$ also
has the same form:
\begin{equation}\label{EqChi2}
\chi^{(1)}=-e^{-iE_{\rm rec}\tau_1 /\hbar}e^{-(\delta x /[2\Delta x])^2}
\end{equation}
with the parameters as defined below Eq.~(\ref{EqChi1}). The interpretation
is the same as below Eq.~(\ref{EqChi1}) and has the same type of scaling.
As in the previous section, if the phase is corrected this gives a value
\begin{equation}\label{EqEps2}
\varepsilon ^{(1)} = \left( \frac{\delta x}{2\Delta x}\right)^2
=\frac{K^2\tau_1^2k_BT_{\rm eff}}{2M}
\end{equation}
where $\tau_1$ is the time between $\pi$ pulses. This expression has the
same form as in the previous section and, thus, has the same scaling.
In most applications,
the separation of the pulses is at least a few times longer than the duration
of the pulses. Since the $\varepsilon^{(1)}$ is proportional to $\tau_1^2$, this
suggests that gates based on excitation followed by a delay and then de-excitation
will have larger decoherence due to photon kick.

To extend the applicability to $\delta t < \tau $,
the derivation from Sec.~\ref{SSSecChi1} can be repeated but with two pulses
with strength $\pi/\delta t$ centered at $t=0$ and $t=\tau_1$. Ignoring
the terms of order $(\delta E\delta t/\hbar)^4$ as in the previous section,
a full derivation gives exactly the same
value as Eq.~(\ref{EqK112}). This can be seen because the two pulses give
a phase accumulation of $-\delta E_i (2\delta t)/2$ and the time between
the pulses gives a phase accumulation of $-\delta E_i (\tau_1 -\delta t)$.
Thus, the result in Eq.~(\ref{EqEps2}) does not depend on the sudden
approximation
and is more accurate than might be expected.

\subsection{One adiabatic pulse}\label{SSSecChiAD}

Instead of exciting the Rydberg state, some gates have a laser pulse that is detuned
so that the population adiabatically evolves, $\sim 100$\%
of the population is in state $|1\rangle$ at the end of the pulse. This is
apparently quite different from the previous two cases, because the
admixture of Rydberg state can be small and ``virtual" when the detuning
is large. However, the derivation
below shows that the same form for $\varepsilon$ results. 

For this case, there is a detuning, $\Delta$ of the laser from $\hbar\omega_{R1}$
so the Eq.~(\ref{eqSchkd}) is modified to
\begin{eqnarray}
i\hbar\frac{\partial\phi_{1,i}(t)}{\partial
t}&=&\left(\bar{E}_i+\frac{\hbar\Delta-\delta E}{2}\right)\phi_{1,i}(t)+
\frac{\hbar\Omega (t)}{2}\phi_{R,i}(t)\nonumber\\
i\hbar\frac{\partial\phi_{R,i}(t)}{\partial
t}&=&\left(\bar{E}_i-\frac{\hbar\Delta-\delta E}{2}\right)\phi_{R,i}(t)+
\frac{\hbar\Omega (t)}{2}\phi_{1,i}(t)\nonumber\\
\end{eqnarray}
Because the gate is adiabatic, the final probability in $|1\rangle$
is $\simeq 1$ so
only the difference in the phase accumulated in $\phi_{1,i}$ compared
to $\phi_{0,i}$ is important. This will lead to
\begin{equation}
{\cal K}_{11}={\cal K}_{11}(\delta E=0)e^{-i\alpha }.
\end{equation}
If $\alpha$ can be written in the form $\alpha =\delta E\tau_a /\hbar$,
then this example will have the same form and interpretation as the
previous two sections. To calculate $\alpha$, we integrate the time
dependent difference between the adiabatic energy and that with $\delta E=0$
and the phase accumulated in $\phi_{0,i}$:
\begin{eqnarray}\label{EqAlph}
\alpha =\frac{1}{\hbar}&\int_{-\infty}^\infty[
&\frac{sgn(\Delta )}{2}\sqrt{(\hbar\Delta -
\delta E)^2+\hbar^2\Omega^2}+\bar{E}-\nonumber\\
&\null & \frac{\hbar sgn(\Delta )}{2}\sqrt{\Delta^2+\Omega^2}
- (\bar{E}-\frac{\delta E}{2}) ] dt
\end{eqnarray}
where $sgn(\Delta )$ means take the sign of $\Delta $. 
The first line is the adiabatic energy of $\phi_{1,i}$, the first
term on the second line is the adiabatic energy when $\delta E=0$,
and the term in parenthesis on the second line is $E(k_i)$. Taylor series
expanding Eq.~(\ref{EqAlph}) to first order in $\delta E$ gives
\begin{equation}
\alpha =\frac{\delta E}{\hbar}\left( \frac{1}{2}\int_{-\infty }^\infty
-\frac{|\Delta |}{\sqrt{\Delta^2 +\hbar^2\Omega^2(t)}}+1dt\right)
\equiv \frac{\delta E\; \tau_a}{\hbar }
\end{equation}
where the term in parenthesis can be identified as 
the time $\tau_a$. By finding
the time dependent eigenstates, the term in parenthesis is the integral of
the probability to be in the Rydberg state for $\delta E=0$:
\begin{equation}\label{EqTau}
\tau_a = \int_{-\infty }^\infty 
\frac{\langle\phi_{R,i}(t)|\phi_{R,i}(t)\rangle|_{\delta E=0} }{\langle\phi_{1,i}(-\infty)|\phi_{1,i}(-\infty)\rangle } dt.
\end{equation}

This leads to a ${\cal K}_{11}^{(ad)}$ with
the same form as Eq.~(\ref{EqK111}) which implies the $\chi$ also
has the same form
with the parameters as defined below Eq.~(\ref{EqChi1}). The interpretation
is the same as below Eq.~(\ref{EqChi1}) and has the same type of scaling.
As in the previous sections, if the phase is corrected this gives a value
\begin{equation}\label{EqEpsAd}
\varepsilon ^{(ad)}(\tau_a ) = \left( \frac{\delta x}{2\Delta x}\right)^2
=\frac{K^2\tau_a^2k_BT_{\rm eff}}{2M}
\end{equation}
where $\tau_a$ is defined in Eq.~(\ref{EqTau}).

For the gate parameters in
Table~\ref{TabAdi},
we numerically found that Eq.~(\ref{EqEpsAd}) accurately reproduced
the results of the full one atom simulations that included vibrational
states.

\subsection{STIRAP pulses}\label{SSSecChiS}

Another method for exciting the Rydberg state is to use
stimulated Raman adiabatic passage (STIRAP). This involves
two photon excitation with an intermediate state. The two
laser pulses only partially overlap and the ordering of the
pulses is typically counterintuitive with the laser coupling
the intermediate state to the Rydberg state coming before the
laser coupling the intermediate state to $|1\rangle$. In this
case, it is not obvious how the timing of the pulses will affect
the momentum kick to the atom.

To understand this case, we will introduce another state $|p\rangle$
to represent the intermediate state.
The Eq.~(\ref{eqSchkd}) is modified to
\begin{widetext}
\begin{eqnarray}
i\hbar\frac{\partial\phi_{1,i}(t)}{\partial
t}&=&[E(k_i)+\hbar\Delta_1]\phi_{1,i}(t)+
\frac{\hbar\Omega_1 (t)}{2}\phi_{p,i}(t)\nonumber\\
i\hbar\frac{\partial\phi_{p,i}(t)}{\partial
t}&=&E(k_i+K_1)\phi_{p,i}(t)+
\frac{\hbar\Omega_1 (t)}{2}\phi_{1,i}(t)+
\frac{\hbar\Omega_R (t)}{2}\phi_{R,i}(t)\nonumber\\
i\hbar\frac{\partial\phi_{R,i}(t)}{\partial
t}&=&[E(k_i+K_1-K_R)-\hbar\Delta_R]\phi_{R,i}(t)+
\frac{\hbar\Omega_R (t)}{2}\phi_{p,i}(t)
\end{eqnarray}
\end{widetext}
where  $K_j$ is the wave
number of the photon, $\Delta_j$ is the detuning,
and $\Omega_j$ is the laser coupling
of the intermediate state
to the states $|j\rangle =|1\rangle$ or $|R\rangle$.
Note the $-$ sign for $\Delta_R$ is because the Rydberg
state is at higher energy than the intermediate state.

In typical STIRAP, the detunings $\Delta$ are chosen so that
states $|1\rangle$ and $|R\rangle$ are degenerate. If the
net recoil is zero $K_1=K_R$, then
this leads to the same equations as when recoil is {\it not}
taken into account. This is because the dark state only involves
$\phi_1$ and $\phi_R$. Thus, there is no decoherence due to recoil
when equal frequency photons are used in STIRAP.

If STIRAP is used to excite the Rydberg state and then de-excite it,
then this leads to a ${\cal K}_{11}^{(S)}$ with
the same form as Eq.~(\ref{EqK111}), which means the $\chi$ also
has the same form
with the parameters as defined below Eq.~(\ref{EqChi1}). The interpretation
is the same as below Eq.~(\ref{EqChi1}) and has the same type of scaling.
As in the previous sections, if the phase is corrected this gives a value
\begin{equation}
\varepsilon ^{(S)} = \left( \frac{\delta x}{2\Delta x}\right)^2
\end{equation}
with the same scaling as previously discussed. The $\tau$ is
still defined as in Eq.~(\ref{EqTau}) but the $\delta x
=\hbar (K_1-K_R)\tau /M$.

\section{Harmonic trapping, $V_{\rm tr}\neq 0$}\label{SSecHT}

The case where the trapping potential is on during the gate manipulations
can not be solved in the general case. However, the case discussed
in Sec.~\ref{SSSecChi2} can be solved for analytically when the atom
starts in the motional ground state and can be done analytically
for small $\varepsilon$ when the atom is in a thermal distribution.
We will only present the derivation for the thermal distribution
but will give the exact result for the ground state at the end of this
section.

Because the duration of each photon pulse is assumed to be much smaller
than the oscillation period, the projection is given by
\begin{equation}
\chi^{(HO)} =\sum_{n=0}^\infty P_n\langle\psi_n|\left(e^{-iH\tau /\hbar }
\right)^\dagger e^{-iKx}e^{-iH\tau /\hbar }e^{iKx}|\psi_n\rangle
\end{equation}
where $P_n=\exp (-\hbar\omega n/[k_BT])/Z$ and $\sum_n P_n=1$
defines $Z$. Going from right to left, the operators inside the expectation
value come from the photon kick, propagating the result for time
$\tau$, the photon kick in the opposite direction, and the time
propagation of the $\langle\psi_n|$. If $\varepsilon$ is small, then
the terms with $K$ can be Taylor series expanded to give
\begin{eqnarray}
\langle\psi_n |...|\psi_n\rangle &=& 1 -K^2\langle\psi_n|x^2-e^{iE_n\tau /\hbar}
xe^{-iH\tau /\hbar }x|\psi_n\rangle\nonumber \\
&=&1-\frac{\hbar K^2}{M\omega}[(1-\cos\omega\tau)
(n+\frac{1}{2})+\frac{i}{2}\sin\omega\tau ]\nonumber\\
\end{eqnarray}
where the terms linear in $x$ give 0 and
we used raising and lowering operators to obtain the second line.
The sum over $n$ can be done analytically and gives
\begin{equation}\label{EqChiHO}
\chi^{(HO)}=1-i\frac{\hbar K^2}{2M\omega}\sin\omega\tau -
\frac{K^2k_BT_{\rm eff}}{M\omega^2}(1-\cos\omega\tau )
\end{equation}
where $T_{\rm eff}$ is given in Eq.~(\ref{EqTeff}). When evaluating
the $|\chi |$, the imaginary term is proportional to $K^4$ and
should be dropped because we are only keeping to order $K^2$.
This gives
\begin{equation}\label{EqEpsHO}
\varepsilon^{(HO)} = \frac{K^2k_BT_{\rm eff}}{M\omega^2}(1-\cos\omega\tau )
\end{equation}
where $\tau$ is the time between the pulses. By Taylor series expanding
the cosine, Eq.~(\ref{EqEps2}) is obtained. More importantly, it shows
that the fractional error in Eq.~(\ref{EqEps2}) is $(\omega\tau )^2/12$
and gives numerical meaning to whether the gate is fast compared
to the trapping period.
Note that having $\tau$ equal to a period of the trap frequency gives
$\varepsilon^{(HO)}=0$.

For the case where there are two very short kicks separated in time
by $\tau$, the ground state projection for a harmonic oscillator of
angular frequency $\omega$ can be found analytically:
\begin{equation}
|\chi|^{(HO,v=0)} = \exp\left[ -\frac{\hbar^2 K^2}{2M\hbar\omega}
(1-\cos\omega \tau)\right].
\end{equation}
This result matches that in Eq.~(\ref{EqChiHO}) by Taylor series expanding the
exponential and noting that $k_B T_{\rm eff}=\hbar\omega/2$ for $T=0$.

\section{Laser focusing}\label{SSecLasFoc}

This section contains effects that arise from the phase and intensity variation
for a Gaussian beam. As pointed out in Ref.~\cite{Gillen-Christandl2016},
the intensity dependence of a focused laser can lead to gate infidelity. In addition, there is also phase variation that will be shown to
be negligible.
To be consistent with the discussion in the other sections, the light
propagates in the $x$ direction and the focused beam intensity varies   in $y$ and $z$.
In this section, we will also consider the
possibility that the focus is not at the origin but at $x=x_0,y=y_0,z$
where $x_0$ is an axial misalignment and $y_0$ is a transverse misalignment.
We will assume that both the misalignments and spatial extent of the
atomic density distribution are small compared to the waist.
If this limit is not satisfied, then the infidelity will be
large.

The phase variation has a linear term from the Gouy phase that decreases
the momentum kick along the beam axis $K\to K-1/x_R$ where $x_R=\pi w_0^2/\lambda$ is
the Rayleigh range with $w_0$ the waist. Typically, $Kx_R\gg 10$ so the relative change
in the axial momentum can be ignored. In each
counter propagating Gaussian beam, there is
also a spatially cubic term in the phase
\begin{equation}
\Delta\Phi = \pm\left(\frac{x^3}{3x_R^3}+\frac{x[(y-y_0)^2+z^2]}{x_Rw_0^2}\right)
\end{equation}
where the $\pm$ is $+$ for right propagating and $-$ for left propagating,
$w_0$ is the waist,
the first term is from the Gouy phase, and the second term is from the
curvature of phase fronts. If the excitation is by 2 photons,
the effect from the phase will
partially cancel due to the change in sign of the two beams.
To estimate the size of the effect from phase variation,
we will use a 2~$\mu$m waist and note that the spatial
extent for 5~$\mu$K
Cs in a 20~kHz trap is 140~nm. To obtain an idea of the importance,
we compare to the phase accumulated by a plane wave $Kx$. The term from the Gouy
phase has a relative contribution of $\sim 10^{-6}$ (for 459~nm) and
$\sim 10^{-7}$ (for 1038 nm). The term from the phase front curvature has
a relative contribution of $\sim 3\times 10^{-5}$ for each. Thus, this effect
might be worth revisiting if infidelities less than $10^{-4}$ become important.
Since the extent of the atom density distribution scales like the square root of
the temperature, if the temperature were 15~$\mu$K, the relative contribution
would be $3\times$ larger.

The focusing leads to a spatial dependence to the electric field strength
\begin{eqnarray}
&E&(x,y,z)=E_0[1-\eta( x,y,z)]\nonumber \\
&\eta&\equiv \frac{(x-x_0)^2}{2x_R^2}
+\frac{(y-y_0)^2+z^2}{w_0^2}
\end{eqnarray}
where $x_R$ is the Rayleigh range and $w_0$ is the waist for single
photon excitation. For two photon excitation,
$1/x_R^2$ is the sum of the squares of the inverse Rayleigh ranges
and $1/w_0^2$ is the sum of the squares of the inverse waists.
As discussed in Ref.~\cite{Gillen-Christandl2016} and the supplemental material of \cite{Graham2019},
this spatial dependence leads to infidelity. 

We will briefly repeat the
derivation leading to infidelity for the $\pi -2\pi -\pi$ gate.
We will assume that the individual pulses are fast enough to ignore
the spatial evolution during a pulse. After a net $2\pi$-pulse
with fractional error $1-\eta$, the spatial wave function for
the $|1\rangle$ state is
\begin{equation}
\psi_{f,1}=\cos [\pi (1 -\eta )]\psi_{in,1}
\simeq -\left(1-\frac{\pi^2\eta^2}{2}\right)\psi_{\rm in,1}
\end{equation}
which leads to
\begin{equation}
\varepsilon^{(G)} = \frac{\pi^2}{2}\langle \eta^2\rangle .
\end{equation}
For typical parameters  $w_0\ll x_R$ but the trapping potential is often substantially
weaker in the axial direction so we will keep all the terms.
We will assume the temperature
is the same along the $x,y,z$ coordinates and that the trap frequency is
the same in the $y,z$ directions. For this case
\begin{eqnarray}\label{EqEpsG}
\varepsilon^{(G)}& =&\frac{\pi^2}{2}\left[
\frac{3\langle x^2\rangle^2+6\langle x^2\rangle x_0^2+x_0^4}{4x_R^4}\right. \nonumber\\
&\null& +\frac{(\langle x^2\rangle +x_0^2)(2\langle y^2\rangle +y_0^2)}{x_R^2w_0^2}\nonumber \\
&\null& \left. 
+\frac{8\langle y^2\rangle^2+8\langle y^2\rangle y_0^2+y_0^4}{w_0^4}
\right]
\end{eqnarray}
where we have used $\langle y^2\rangle = \langle z^2\rangle
=k_BT_{\rm eff,\perp}/(M\omega_\perp^2)$,
$\langle y^4\rangle = \langle z^4\rangle = 3\langle y^2\rangle ^2$,
and
$\langle x^2\rangle = k_BT_{\rm eff,\parallel}/(M\omega_\parallel^2)$
where the symbols are meant to indicate that the $T_{\rm eff}$
and $\omega$ are
different for $x$ and $y,z$.

This is a complicated expression so it is worthwhile
to note that in many cases the effect on the axial motion from focusing (first two
lines) will be substantially smaller than that from the transverse focusing
(last line). In this case, the expression
\begin{equation}
\varepsilon^{(G)}\simeq \frac{\pi^2}{2}\left(\frac{y_0}{w_0}\right)^4
+ 4\pi^2\left(\frac{y_0}{w_0}\right)^2{\cal D}
+4\pi^2{\cal D}^2
\end{equation}
with ${\cal D}= k_B T_{\rm eff,\perp}/(M\omega_\perp^2w_0^2)$
gives a good approximation of the effect from focusing and makes clearer the
dependence on parameters.

There is one tricky aspect that arises for the $\pi -2\pi -\pi$ gate. The
state $|11\rangle$ has nontrivial time evolution. The first $\pi$-pulse
puts this state into a superposition of $|R1\rangle$ and $|11\rangle$ with
most of the population in $|R1\rangle$. The $2\pi$-pulse mainly changes
the sign of the $|11\rangle$ state. This means the last $\pi$-pulse rotates
almost perfectly opposite the initial $\pi$-pulse so that the final superposition
has $|11\rangle$ to order $\eta^4$. Thus, {\it to the order in this section},
$|11\rangle$ has no decoherence. Only the states $|01\rangle$ and $|10\rangle$
will suffer decoherence from focusing.


%

\end{document}